\documentclass[a4paper,12pt]{article}
\usepackage[utf8]{inputenc}
\usepackage{slashed} 
\usepackage{enumerate}
\usepackage{epsfig} 
\usepackage{float} 
\usepackage{subcaption} 
\usepackage{amsmath} 
\usepackage{wasysym} 
\usepackage{mathrsfs} 
\usepackage{amsfonts} 
\usepackage{amsbsy} 
\usepackage{amscd}    
\usepackage{multirow} 
\usepackage{tikz}
\usepackage{slashed}
\usepackage{jheppub}
\usepackage[miktex]{gnuplottex}
\usetikzlibrary{arrows,positioning,shapes.geometric} 
\usepackage{color}
\usepackage{verbatim}
\usepackage{xcolor}
\DeclareUnicodeCharacter{2212}{\ensuremath{-}}
\allowdisplaybreaks

\usepackage{tikz} 
\usetikzlibrary{shapes,arrows,positioning,automata,backgrounds,calc,er,patterns}
\usepackage{tikz-feynman}
\tikzfeynmanset{compat=1.0.0}

\usepackage{soul}

\title{Dark Matter Phenomenology in 2HDMS in light of the 95 GeV excess}
\author[a]{Juhi Dutta,}  
\author[b]{Jayita Lahiri,}
\author[d]{Cheng Li,}
\author[b,c]{Gudrid Moortgat-Pick,}
\author[b]{Sheikh Farah Tabira}
\author[b]{and Julia Anabell Ziegler}
\affiliation[a]{Homer L. Dodge Department of Physics and Astronomy,
University of Oklahoma, Norman, OK 73019, USA}
\affiliation[b]{II. Institut f{\"u}r Theoretische Physik,  Universit{\"a}t Hamburg, Luruper Chaussee 149, 22761 Hamburg, Germany}
\affiliation[c]{Deutsches Elektronen-Synchrotron DESY, 
Notkestr. 85, 22607 Hamburg, Germany}
\affiliation[d]{School of Science, Sun Yat-Sen University, Gongchang Road 66, 518107 Shenzhen, China}
\emailAdd{juhi.dutta@ou.edu}
\emailAdd{jayita.lahiri@desy.de}
\emailAdd{gudrid.moortgat-pick@desy.de}
\emailAdd{cheng.li@desy.de}
\emailAdd{sheikh.farah.tabira@desy.de}
\emailAdd{julia.ziegler@desy.de}
 
\abstract{The Two Higgs Doublet model extended with a complex scalar singlet (2HDMS) is a well-motivated Beyond Standard Model candidate addressing several open problems of nature. In this work, we focus on the dark matter (DM) phenomenology of the complex scalar singlet  where the real part of the complex scalar obtains a vacuum expectation value. The model is characterized by  an enlarged Higgs spectrum comprising six physical Higgs bosons and a pseudoscalar DM candidate. We address the impact of accommodating the 95 GeV excess on the 2HDMS parameter space and DM observables after including all theoretical and experimental constraints. Finally, we look into the prospects of  this scenario at HL-LHC and future lepton colliders for a representative benchmark.}

\preprint{DESY-23-114}

\begin{document}

\maketitle

\section{Introduction}
The evidence for dark matter (DM) has been strongly established from experimental observations involving gravitational interactions such as the rotational velocity of galaxies, the Bullet Cluster~\cite{Barrena:2002dp} and from  the Cosmic Microwave Background (CMB)~\cite{Planck:2018vyg}. However, the Standard Model (SM) of Particle Physics does not provide a suitable candidate for cold DM, therefore, it is compelling to venture beyond the Standard Model (BSM) with a plethora of possible candidates for DM, varying from scalar-, fermion- or vector-like new particles, extending the SM and have been the subject of intense  scrutiny for decades. \\
The discovery of the 125 GeV Higgs~\cite{ATLAS:2012yve,CMS:2012qbp} during Run 1 at the Large Hadron Collider (LHC) in 2012 paved the way for the success of the SM, but establishing it as a complete model on its own can not be done. There still remains a question on whether the Higgs observed is pure SM-like or a signature of a larger BSM Higgs sector. Ongoing precise experimental measurements at the LHC~\cite{ATLAS:2022vkf, CMS:2022dwd} still allow the  possibility to accommodate Beyond Standard Model effects  in the Higgs couplings. It is therefore mandatory to focus also on high precision studies of the Higgs and electroweak sector at a linear collider, complementing the searches for BSM physics at the high luminosity LHC upgrade  (HL-LHC). Furthermore, a new excess has been observed  both at the former Large Electron-Positron Collider (LEP) in the $b\bar{b}$ mode~\cite{LEPWorkingGroupforHiggsbosonsearches:2003ing} as well as at the $\gamma\gamma$~\cite{CMS:2017yta,ATLAS:2018xad,CMS:2023yay} and $\tau\tau$ modes~\cite{CMS:2022rbd} at the LHC experiment CMS  (and more recently in the $\gamma\gamma$ mode at the LHC experiment ATLAS\footnote{During the conclusion of our study, a new result at ATLAS showing a slight excess ($\sim 1.7 \sigma$) was observed in the $\gamma\gamma$ channel \cite{ATLAS-CONF-2023-035} and has been recently studied in Ref. \cite{Biekotter:2023oen} in S2HDM (i.e. the U(1) symmetric 2HDMS). For our study, however, we have only considered the LEP and CMS excesses.}) at a mass of around 95 GeV.  \\
In the light of these results, several BSM models address such a light scalar excess, e.g.\ in extended Higgs models, including multi-Higgs models such as 2HDM \cite{Fox:2017uwr, Haisch:2017gql,  Benbrik:2022azi, Benbrik:2022dja, Azevedo:2023zkg, Belyaev:2023xnv}, N2HDM \cite{Biekotter:2019kde}, UN2HDM \cite{Aguilar-Saavedra:2023vpd, Banik:2023ecr}, 2HDMS \cite{Heinemeyer:2021msz}, S2HDM \cite{Biekotter:2021ovi, Biekotter:2023jld, Biekotter:2023oen} and supersymmetric extensions such as NMSSM \cite{Cao:2016uwt, Cao:2019ofo, Choi:2019yrv, Biekotter:2021qbc, Li:2022etb} (see Ref.~\cite{PhysRevD.100.035023, Biekotter:2017xmf, Liu:2018ryo, Liu:2018xsw, Biekotter:2019gtq, Richard:2017kot, Aguilar-Saavedra:2020wrj, Biekotter:2020cjs, Coloretti:2023wng, Bhattacharya:2023lmu, Ashanujjaman:2023etj, Escribano:2023hxj} for more details on other allowed models accommodating the 95 GeV excess). It has been shown that such an excess can be fit in the N2HDM and 2HDMS for the Type II 2HDM Higgs sector. In addition, the $\tau\tau$ excess can be fit in the Type IV N2HDM \cite{Biekotter:2022jyr}. \\
In this study, we investigate the Type II Two Higgs Doublet model augmented with a complex scalar singlet (2HDMS) in the context of the 95 GeV excess  in the $\gamma\gamma$ and $b\bar{b}$ modes, observed at CMS and LEP respectively, in conjunction with DM. We consider the case where the singlet scalar develops a vacuum expectation value (\textit{vev}) resulting in mixing with the 2HDM Higgs sector leading to three CP-even scalar Higgs $h_1,h_2,h_3$, (one of which must be the SM-like 125 GeV Higgs), one CP odd pseudoscalar A, a pair of charged Higgses $H^{\pm}$ and a pseudoscalar DM candidate $A_S$. The presence of an extra singlet-dominated CP-even scalar compared to the 2HDM~\cite{Branco:2011iw,Campos:2017dgc} provides the possibility of accommodating a light 95 GeV scalar in order to explain the $\gamma\gamma$ and $b\bar{b}$ excesses. Furthermore, the presence of a DM candidate leads to missing energy signatures at collider experiments, such as mono-X + missing energy searches  (where X = jet, Z, W, $\gamma$) at LHC. In the presence of extended Higgs sectors, the gluon fusion and vector boson fusion channels can lead to mono-jet + missing energy and two forward jets + missing energy signatures. On the other hand, at lepton colliders such as ILC, CLIC and muon colliders, mono-$\gamma,Z$ + missing energy channels are relevant signatures for dark matter searches. In our model, we can embed the 95 GeV excess while being consistent with both theoretical as well as experimental constraints and identify the relevant parameter space. We choose a representative benchmark point from there and perform a scan over the parameter space to explore the DM phenomenology  and prospects of 2HDMS at the high luminosity LHC (HL-LHC) for gluon fusion, vector boson fusion  and mono-$\gamma,Z$ channels for future lepton colliders such as  ILC and muon colliders.  Thus, the salient points of our work are:
\begin{itemize}
    \item An extension of Type II 2HDM with a complex singlet scalar (2HDMS) is considered in the light of the excess observed at 95 GeV from current observations at CMS and LEP. Alongside accommodating the 95 GeV excess, this model can also provide a viable dark matter candidate.

     \item We have  derived the boundedness-from-below (bfb) conditions  for the model and observe that it provides the most stringent constraints on the parameter space among the theoretical constraints. 
     
    \item We have scanned over the parameter space and chosen a representative benchmark satisfying all theoretical and experimental constraints. We observed that the parameters  $\delta^{\prime}_{25}$  (the effective DM-Higgs portal coupling), $\tan \beta$ and $m^{2\prime}_S$ (coefficient of the DM mass term in the Lagrangian) are stringently constrained from the current data from direct detection searches.      
   
     \item We discussed the prospects of observing signals at HL-LHC in the mono-jet and 2j + $\slashed{E}_T$ channels. While the invisible branching ratio in the benchmark scenario is $\sim 25\%$, owing to the heavy Higgs mass, prospects of observing these signals at LHC are weak using cut-and-count analyses.  We also discuss some potentially important signal processes, such as mono-$\gamma$ and mono-Z + missing energy, at future lepton colliders and highlight the advantage of a $\mu^+\mu^-$ collider in probing such a scenario.
\end{itemize} 
The paper is organized as follows: in sec.~\ref{sec:model} we introduce the model followed by a discussion on the relevant theoretical and experimental constraints in sec.~\ref{sec:const}. In sec.~\ref{sec:dm} we present the allowed parameter space regions subjected to theoretical and experimental constraints and discuss the prospects of observing 2HDMS at present and future hadron and lepton colliders in sec.~\ref{sec:collider}. We summarise our results in sec.~\ref{sec:summ}.

\section{The Model}
\label{sec:model}
In this work,  we consider the CP-conserving Type II Two Higgs Doublet Model (2HDM) augmented with a complex scalar singlet (2HDMS)~\cite{Baum:2018zhf} with a softly broken $Z_2$-2HDM sector consistent with the flavour changing neutral currents (FCNCs) with an additional broken $Z^{\prime}_2$ symmetry for the complex singlet. In previous works several different symmetries have been considered for the singlet sector phenomenology including $U(1)$~\cite{Biekotter:2021ovi, Biekotter:2023jld}, 
$Z_3$~\cite{Baum:2018zhf,Heinemeyer:2021msz} and $Z^{\prime}_2$~\cite{Dutta:2022xbd} as well as CP-violation studies~\cite{Muhlleitner:2021cci}. For conserved $Z^{\prime}_2$ the complex scalar singlet  does not develop a vacuum expectation value (\textit{vev}), i.e. $<S>=0$  and is stabilized under the new $Z^{\prime}_2$-symmetry, i.e.   $S$ is odd under $Z_2^{\prime}$ while the SM fields are even under it. In this work, we consider the case where the $Z^{\prime}_2$ is broken and the real part of the complex scalar develops a \textit{vev} and mixes with the Higgs sector while the imaginary part of the scalar is postulated to be odd under the $Z^{\prime}_2$-symmetry and constitutes either the full or at least part of the DM content of nature.  
The symmetries of the model are summarised in Table.~\ref{tab:2HDMS_symmetries}. The scalar fields $\Phi_1$ and $S$ are even under $Z_2$ while $\Phi_2$ is odd under $Z_2$. On the other hand, $\Phi_1$ and $\Phi_2$ are even under the new symmetry $Z^{\prime}_2$ while $S$ is odd under $Z^{\prime}_2$. The $Z^{\prime}_2$  also breaks dynamically such that the real part of the scalar mixes with the Higgs bosons while the imaginary part constitutes the DM candidate.  
\begin{table}[h]
\scriptsize
    \centering
    \begin{tabular}{|c|c|c|}
        \hline
        Symmetry & Transformation & Effect \\
        \hline
        $V_{2HDM}$ symmetric under $U(1)$, & $\Phi_j \overset{U(1)}{\rightarrow} e^{i\theta}\Phi_j, \quad \Phi_j^\dagger \overset{U(1)}{\rightarrow} e^{-i\theta}\Phi_j^\dagger$ & 2HDM potential \\
        all parameters real & & symmetric under CP \\
        \hline 
        $V_{2HDM}$ symmetric under $Z_2$ & $\Phi_1 \overset{Z_2}{\rightarrow}-\Phi_1, \quad \Phi_2 \overset{Z_2}{\rightarrow}\Phi_2$ & avoids FCNC \\
        (softly broken by parameter $m_{12}^2 \neq 0$) & & \\
        (spontaneously broken by vevs $v_1,v_2$) & & \\
        \hline 
        $V_{2HDMS}$ symmetric under $Z^{\prime}_2$ & $\Phi_j \overset{Z^{\prime}_2}{\rightarrow}\Phi_j, \quad S\overset{Z^{\prime}_2}{\rightarrow}-S$ & stabilization of DM\\
        (spontaneously broken by vev $v_S$) & & \\
        \hline
    \end{tabular}
    \caption{Symmetries of the 2HDMS scalar potential  and their effects on the different scalar fields.}
    \label{tab:2HDMS_symmetries}
\end{table} 
Therefore, the scalar potential $V$ follows, 
\begin{equation}
\label{eq:sp}
 V = V_\mathrm{2HDM}+V_S,
\end{equation}
where the softly broken $Z_2$-symmetric 2HDM potential is,
\begin{align}
  V_\mathrm{2HDM} &= m^2_{11} \Phi_1^{\dagger}\Phi_1 + m^2_{22} \Phi_2^{\dagger}\Phi_2-(m^2_{12} \Phi_1^{\dagger}\Phi_2+h.c.) \nonumber \\
  & \quad + \frac{\lambda_1}{2} (\Phi_1^{\dagger}\Phi_1)^2  + \frac{\lambda_2}{2} (\Phi_2^{\dagger}\Phi_2)^2 + \lambda_3 (\Phi_1^{\dagger}\Phi_1) (\Phi_2^{\dagger}\Phi_2) \nonumber \\
  & \quad + \lambda_4 (\Phi_1^{\dagger}\Phi_2) 
  (\Phi_2^{\dagger}\Phi_1) + [\frac{\lambda_5}{2} (\Phi_1^{\dagger}\Phi_2)^2+h.c.],
\end{align}
and the singlet potential $V_S$ is,
\begin{align}
  V_S  &= m^2_S S^{*}S + (\frac{m^{\prime 2}_S}{2} S^2 + h.c.) \nonumber \\  
  & \quad + (\frac{\lambda^{\prime\prime}_1}{24}S^4 +h.c.)+(\frac{\lambda_2^{\prime\prime}}{6}(S^2 S^{*}S) +h.c.) +\frac{\lambda_3^{\prime\prime}}{4} (S^{*}S)^2  \nonumber \\
  &\quad  + S^{*}S[\lambda_1^{\prime} \Phi_1^{\dagger}\Phi_1 + 
  \lambda_2^{\prime}\Phi_2^{\dagger}\Phi_2]+  [S^2 (\lambda^{\prime}_4 \Phi_1^{\dagger}\Phi_1+\lambda_5^{\prime}\Phi_2^{\dagger}\Phi_2) +h.c.].
\end{align}
\newpage 
\noindent Therefore, the full scalar potential is,
\begin{align}
\label{eq:cspot}
   V &= m^2_{11} \Phi_1^{\dagger}\Phi_1 + m^2_{22} \Phi_2^{\dagger}\Phi_2-(m^2_{12} \Phi_1^{\dagger}\Phi_2+h.c.)  \nonumber \\ 
   & \quad + \frac{\lambda_1}{2} (\Phi_1^{\dagger}\Phi_1)^2 +  \frac{\lambda_2}{2} (\Phi_2^{\dagger}\Phi_2)^2 + \lambda_3 (\Phi_1^{\dagger}\Phi_1) (\Phi_2^{\dagger}\Phi_2) \nonumber \\
   & \quad + \lambda_4 (\Phi_1^{\dagger}\Phi_2) (\Phi_2^{\dagger}\Phi_1) + [\frac{\lambda_5}{2} (\Phi_1^{\dagger}\Phi_2)^2+h.c.] \nonumber \\
   & \quad + m^2_S S^{*}S + (\frac{m^{\prime2}_S}{2} S^2 + h.c.) \nonumber \\
   & \quad  + (\frac{\lambda^{\prime\prime}_1}{24}S^4 +h.c.)+(\frac{\lambda_2^{\prime\prime}}{6}(S^2 S^{*}S) +h.c.)+\frac{\lambda_3^{\prime\prime}}{4} (S^{*}S)^2 + \nonumber \\
   & \quad S^{*}S[\lambda_1^{\prime} \Phi_1^{\dagger}\Phi_1 + 
   \lambda_2^{\prime}\Phi_2^{\dagger}\Phi_2] +  [S^2 (\lambda^{\prime}_4 \Phi_1^{\dagger}\Phi_1+\lambda_5^{\prime}\Phi_2^{\dagger}\Phi_2) +h.c.],
\end{align}
where,
\begin{equation}
\Phi_i = 
\begin{pmatrix}
  \phi^{\pm}_i  \\
  \frac{1}{\sqrt{2}}(v_i+h_i+ i a_i)\\
 \end{pmatrix} , \quad  i=1,2, 
\end{equation}
are the Higgs doublets while the complex scalar singlet is,
\begin{equation}
 S = \frac{1}{\sqrt{2}}(v_S+h_S + i a_S).
\end{equation}
The vacuum expectation value(\textit{vev}) of the Higgs doublets and complex scalar singlet are denoted by $v_1, v_2$ and $v_S$ respectively.
The minimization conditions of the scalar potential are,
\begin{eqnarray}
      m_{11}^2 v_1 -m_{12}^2 v_2 + \frac{\lambda_1}{2} v_1^3 + \frac{\lambda_{345}}{2} v_1 v_2^2 + (\frac{\lambda_1'}{2}v_1 + \lambda_4' v_1)v_S^2   &= 0, \\
     m_{22}^2 v_2 -m_{12}^2 v_1 + \frac{\lambda_2}{2} v_2^3 + \frac{\lambda_{345}}{2} v_1^2 v_2 + (\frac{\lambda_2'}{2} v_2 + \lambda_5' v_2) v_S^2&=0,\\
     m_S^2 v_S + m_S'^2 v_S + \frac{\lambda_1''}{12} v_S^3 + \frac{\lambda_2''}{3} v_S^3 + \frac{\lambda_3''}{4} v_S^3  + \frac{v_S}{2}(\lambda_1' v_1^2 + \lambda_2' v_2^2) + \nonumber \\  v_S(\lambda_4' v_1^2 + \lambda_5' v_2^2)&=0.
\end{eqnarray}
After electroweak symmetry breaking (EWSB), $m^2_{11}, m^2_{22}$ and $m^2_S$ are replaced by the mi\-ni\-mi\-zation equations, thereby reducing the total  free parameters in the theory to the following,
\begin{equation}
  \lambda_1, \lambda_2,  \lambda_3,  \lambda_4,  \lambda_5, m^2_{12}, \tan \beta, v_S,
    m^{2\prime}_S, \lambda^{\prime}_1, \lambda^{\prime}_2, \lambda^{\prime}_4, \lambda^{\prime}_5, \lambda^{\prime\prime}_1 , \lambda^{\prime\prime}_2, \lambda^{\prime\prime}_3. 
  \end{equation}
For simplicity, we  choose to set the quartic couplings,
\begin{equation}
 \lambda^{\prime\prime}_1=\lambda^{\prime\prime}_2. \,\,
\end{equation}
 Such a choice of these quartic couplings may  affect the DM mass and couplings, as seen in eq.~\ref{masseq}-\ref{eq:DM_couplings_15dof_mass_basis_quatr} and affect both dark matter and collider phenomenology which we do not consider in this work. 
 
\subsection{Higgs Sector}
The Higgs sector in the 2HDMS consists of three CP-even Higgs bosons $h_1,h_2,h_3$, one CP-odd pseudoscalar $A$ and a pair of charged Higgs bosons $H^{\pm}$. Of the three CP-even Higgs bosons, one of them is consistent with the 125 GeV SM-like Higgs observed experimentally at the  LHC, 
a necessary condition for any BSM model. The scalar mass matrix is,
\begin{equation}
M^2_{S}=  
 \begin{pmatrix}
   m^2_{12}\frac{v_2}{v_1}+\lambda_1v^2_1& -m^2_{12}+\lambda_{345}v_1v_2 & (\lambda^{\prime}_1 +   2\lambda^{\prime}_4)v_1v_S\\
   -m^2_{12}+\lambda_{345}v_1v_2& m^2_{12}\frac{v_1}{v_2}+\lambda_2v^2_2 & (\lambda^{\prime}_2 +   2\lambda^{\prime}_5)v_2v_S\\
 (\lambda^{\prime}_1 +   2\lambda^{\prime}_4)v_1v_S& (\lambda^{\prime}_2 +   2\lambda^{\prime}_5)v_2v_S& (\frac{5\lambda^{\prime\prime}_1}{6} + \frac{\lambda^{\prime\prime}_3}{2})v^2_S\\
 \end{pmatrix}.
\end{equation}
For the pseudoscalar and the charged Higgs sectors, the Goldstone modes are absorbed by the $W$ and $Z$ bosons after electroweak symmetry breaking. The charged Higgs sector remains the same as in the 2HDM.   
In this study, we assume the lightest CP-even Higgs, $h_1$ to be the 95 GeV excess observed at LEP in the $b\bar{b}$ mode ~\cite{LEPWorkingGroupforHiggsbosonsearches:2003ing} and at CMS in the $\gamma\gamma$ mode~\cite{CMS-PAS-HIG-20-002} while the second CP-even Higgs is set as the SM-like 125 GeV Higgs.  
\subsection*{Mass Basis}
The Higgs sector of this model consists of three CP-even Higgs bosons, i.e, $h_1$, $h_2$, $h_3$ with singlet-doublet mixing characterized by $\alpha_1, \alpha_2, \alpha_3$. The pseudoscalar Higgs and the pseudoscalar component of the singlet do not mix due to the applied symmetries. Therefore the particle content of the model consists of $h_1, h_2, h_3, A, H^{\pm}$ and $A_S$.
In the CP-conserving case, the 15 remaining free parameters of the model are shown in eq.~\ref{eq:freepar_inte}:
\begin{equation}
    \lambda_1, \lambda_2,  \lambda_3,  \lambda_4,  \lambda_5, m^2_{12}, \tan \beta, v_S,
    m^{2\prime}_S, \lambda^{\prime}_1, \lambda^{\prime}_2, \lambda^{\prime}_4, \lambda^{\prime}_5, \lambda^{\prime\prime}_1 = \lambda^{\prime\prime}_2, \lambda^{\prime\prime}_3.
\label{eq:freepar_inte}
\end{equation}
Fixing the mass basis one derives the corresponding free parameters as shown in eq.~\ref{eq:freepar_mass}:
\begin{equation}
    \begin{split}
        m_{h_1}, m_{h_2}, &m_{h_3}, m_A, m_{A_S}, m_{H^\pm}, \delta_{14}'=\lambda^{\prime}_4 - \lambda^{\prime}_1, \delta_{25}'=\lambda^{\prime}_5 - \lambda^{\prime}_2, \\
        &\tan \beta, v_S, c_{h_1 bb}, c_{h_1 tt}, \Tilde{\mu}^2, m_{S}'^2, \mathrm{alignm}. 
    \end{split}
\label{eq:freepar_mass}
\end{equation} 
where  $\tilde{\mu}^2$ and $\mathrm{alignm}$ is defined in eq.~\ref{eq:massbasis} and eq.~\ref{eq:alignm} respectively.
The couplings in the scalar potential are then rewritten in terms of the mass basis parameters. The relation between interaction basis and mass basis parameters is shown below in eq.~\ref{eq:massbasis}.  A few comments are in order regarding our choice of parameters in the mass basis. A closer inspection of eq.~\ref{eq:cspot}, reveals that the couplings of DM particle $A_S$ (which is essentially same as the imaginary part of the complex scalar field $S$), to all the neutral scalars ($h_i$) will involve the particular combinations $\lambda^{\prime}_4 - \lambda^{\prime}_1$ and $\lambda^{\prime}_5 - \lambda^{\prime}_2$, of the portal couplings. Since we are particularly interested in the DM phenomenology, these combinations will play a crucial role in our study, as will be clear in the upcoming sections. Keeping this in mind, we use these combinations, as free parameters in the mass basis, naming them $\delta_{14}'$ and $\delta_{25}'$ respectively. 
\begin{align}\label{eq:massbasis}
   \lambda_1 &=\frac{1}{v^2\cos^2\beta}(\Sigma^3_{i=1} m^2_i R^2_{i1}-\Tilde{\mu}^2\sin^2\beta), \nonumber \\
   \lambda_2 &= \frac{1}{v^2\sin^2\beta}(\Sigma^3_{i=1} m^2_i R^2_{i2}-\Tilde{\mu}^2\cos^2\beta),\nonumber \\
   \lambda_3 &= \frac{1}{v^2}(\frac{1}{\sin\beta\cos\beta}\Sigma^3_{i=1} m^2_i R_{i1}R_{i2}-\Tilde{\mu}^2+2m^2_{H^{\pm}}),\nonumber \\
   \lambda_4 &= \frac{1}{v^2}(m^2_A+\Tilde{\mu}^2-2m^2_{H^{\pm}}),\nonumber \\
   \lambda_5 &= \frac{1}{v^2}(-m^2_A+\Tilde{\mu}^2), \nonumber \\
   \lambda^{\prime}_1&=\frac{1}{3} (\frac{1}{vv_S\cos\beta}\Sigma^3_{i=1} m^2_iR_{i1}R_{i3} - 2\delta_{14}'),\nonumber \\
   \lambda^{\prime}_2&=\frac{1}{3}(\frac{1}{vv_S\sin\beta}\Sigma^3_{i=1} m^2_iR_{i2}R_{i3} - 2\delta_{25}'),\nonumber \\
   \lambda^{\prime}_4&= \frac{1}{3} (\frac{1}{vv_S\cos\beta}\Sigma^3_{i=1} m^2_iR_{i1}R_{i3} + \delta_{14}'), \nonumber \\
   \lambda^{\prime}_5&= \frac{1}{3}(\frac{1}{vv_S\sin\beta}\Sigma^3_{i=1} m^2_iR_{i2}R_{i3} + \delta_{25}'), \nonumber \\
   \lambda_1^{{\prime}{\prime}} &= \lambda_2^{{\prime}{\prime}} = - \frac{3}{2 v_S^2} (2 m^{2\prime}_S +2 v^2(\frac{1}{3}(\frac{1}{vv_S\cos\beta}\Sigma^3_{i=1} m^2_iR_{i1}R_{i3} +\delta_{14}')\cos^2\beta \nonumber \\
   & \quad + \frac{1}{3}(\frac{1}{vv_S\sin\beta}\Sigma^3_{i=1} m^2_iR_{i2}R_{i3} +\delta_{25}')\sin^2\beta) + m_{A_S}^2), \nonumber \\
   \lambda_3^{{\prime}{\prime}} &= \frac{1}{3}(\frac{6}{v_S^2} \Sigma_{i=1}^{3} m_i^2 R_{i3}^2 \nonumber \\
   & \quad +\frac{15}{2 v_S^2} (2 m^{2\prime}_S +2 v^2(\frac{1}{3}(\frac{1}{vv_S\cos\beta}\Sigma^3_{i=1} m^2_iR_{i1}R_{i3} +\delta_{14}')\cos^2\beta \nonumber \\
   & \quad + \frac{1}{3}(\frac{1}{vv_S\sin\beta}\Sigma^3_{i=1} m^2_iR_{i2}R_{i3} +\delta_{25}')\sin^2\beta) + m_{A_S}^2)), \nonumber \\
   m^2_{12} &= \Tilde{\mu}^2 \cdot \sin\beta\cos\beta,
\end{align}
where $R_{ij}$ are the elements of the rotation matrix $R$ in the CP-even Higgs sector defined as,
\begin{equation}
 R = 
 \begin{pmatrix}
 \scriptsize
  c_{\alpha_1}  c_{\alpha_2}  & s_{\alpha_1} c_{\alpha_2} & s_{\alpha_2} \\
  -s_{\alpha_1} c_{\alpha_3} - c_{\alpha_1} s_{\alpha_2} s_{\alpha_3} & c_{\alpha_1} c_{\alpha_3} - s_{\alpha_1} s_{\alpha_2} s_{\alpha_3}& c_{\alpha_2} s_{\alpha_3}\\
  s_{\alpha_1} s_{\alpha_3} - c_{\alpha_1} s_{\alpha_2} c_{\alpha_3} & - c_{\alpha_1} s_{\alpha_3 }- s_{\alpha_1} s_{\alpha_2} c_{\alpha_3} &  c_{\alpha_2} c_{\alpha_3}\\ 
 \end{pmatrix} ,
\end{equation}
where $s_\alpha$ denotes $\sin(\alpha)$ and $c_\alpha$ denotes $\cos(\alpha)$. \\
In addition, we replace the scalar mixing angles $\alpha_i$, $i=1,2,3$ by using  reduced couplings defined as,
\begin{align}
    c_{h_1 t t}=\frac{\sin(\alpha_1)\cos(\alpha_2)}{\sin(\beta)}, \\
    c_{h_1 b b}=\frac{\cos(\alpha_1)\cos(\alpha_2)}{\cos(\beta)},
\end{align} 
and the alignment condition~\cite{Heinemeyer:2021msz} i.e. 
\begin{align}
    \alpha_3 = \frac{\beta - \alpha_1 - \arcsin(\mathrm{alignm})}{\text{sgn}(\alpha_2)} \approx \frac{\beta - \alpha_1 - \pi/2}{\text{sgn}(\alpha_2)}, \\
    \Rightarrow \mathrm{alignm}=|\sin(\beta - (\alpha_1 + \alpha_3 \cdot \text{sgn}(\alpha_2)))| \approx 1 .
\label{eq:alignm}
\end{align} 
\subsection{Dark Sector}
After EWSB, $A_S$ constitutes the pseudoscalar DM candidate. The squared mass of the DM candidate is, 
\begin{align}
    m_{A_S}^2  &= -(2 m_S'^2 + \frac{2 \lambda_1''}{3} v_S^2 +2(\lambda_4' v_1^2 + \lambda_5' v_2^2)) .
    \label{masseq}
\end{align}
The DM couples to the SM particles via the CP-even Higgs bosons. The trilinear and quatrilinear couplings of the DM candidate to scalar Higgs particles can be written as:
\begin{align}
    \frac{\lambda_{h_j A_S A_S}}{v} 
    &= - [(\lambda_1' - 2\lambda_4')c_{\beta} R_{j1} + (\lambda_2' - 2\lambda_5')s_{\beta} R_{j2} - \frac{v_S}{2v}(\lambda_1'' - \lambda_3'')R_{j3}] ,  \label{eq:DM_couplings_tril} \\
    \lambda_{h_j h_k A_S A_S} 
    &= - [(\lambda_1' - 2\lambda_4')R_{j1}R_{k1} + (\lambda_2' - 2\lambda_5')R_{j2}R_{k2} - \frac{1}{2}(\lambda_1'' - \lambda_3'')R_{j3}R_{k3}] , \label{eq:DM_couplings_quatr}
\end{align}
where $R$ is the scalar rotation matrix.  
In the 15 degree-of-freedom (d.o.f) mass basis the couplings can be written as:
\begin{subequations}\label{eq:DM_couplings_15dof_mass_basis}
\begin{align}
    \frac{\lambda_{h_j A_S A_S}}{v} 
    &=[\frac{\sum_{i=1}^{3}m_{h_i}^2 R_{i1}R_{i3}}{3v v_S \cos(\beta)} + \frac{4\delta_{14}'}{3}]c_{\beta}R_{j1} \nonumber \\ 
    &\quad + [\frac{\sum_{i=1}^{3} m_{h_i}^2 R_{i2}R_{i3}}{3v v_S \sin(\beta)} + \frac{4\delta_{25}'}{3}]s_{\beta}R_{j2} \nonumber \\ 
    & \quad - [\frac{2}{v v_S}(2m_S'^2 + m_{A_S}^2+(\frac{\sum_{i=1}^{3}m_{h_i}^2 R_{i1}R_{i3}}{3v v_S \cos(\beta)} + \frac{\delta_{14}'}{3})2v^2c_{\beta}^2 \nonumber \\ 
    & \quad \quad + (\frac{\sum_{i=1}^{3} m_{h_i}^2 R_{i2}R_{i3}}{3v v_S \sin(\beta)} + \frac{\delta_{25}'}{3})2v^2s_{\beta}^2) + \frac{\sum_{i=1}^{3} m_i^2 R_{i3}^2}{v v_S}]R_{j3},   
\end{align}
\end{subequations}
\newpage 
\begin{subequations}
\begin{align}
    \lambda_{h_j h_k A_S A_S}
    &=[\frac{\sum_{i=1}^{3}m_{h_i}^2 R_{i1}R_{i3}}{3v v_S \cos(\beta)} + \frac{4\delta_{14}'}{3}]R_{j1}R_{k1} \nonumber \\ 
    &\quad + [\frac{\sum_{i=1}^{3} m_{h_i}^2 R_{i2}R_{i3}}{3v v_S \sin(\beta)} + \frac{4\delta_{25}'}{3}]R_{j2}R_{k2} \nonumber \\ 
    & \quad - [\frac{2}{v_S^2}(2m_S'^2 + m_{A_S}^2+(\frac{\sum_{i=1}^{3}m_{h_i}^2 R_{i1}R_{i3}}{3v v_S \cos(\beta)} + \frac{\delta_{14}'}{3})2v^2c_{\beta}^2 \nonumber \\ 
    & \quad \quad + (\frac{\sum_{i=1}^{3} m_{h_i}^2 R_{i2}R_{i3}}{3v v_S \sin(\beta)} + \frac{\delta_{25}'}{3})2v^2s_{\beta}^2) + \frac{\sum_{i=1}^{3} m_i^2 R_{i3}^2}{v_S^2}]R_{j3}R_{k3}.  
    \label{eq:DM_couplings_15dof_mass_basis_quatr}
\end{align}
\end{subequations}
These couplings influence the DM observables, namely direct detection DM-proton and DM-neutron cross-section, indirect detection DM annihilation cross-section and relic density. The main process contributing to the direct detection cross-section is elastic scattering via the exchange of a CP-even Higgs boson and is shown in the appendix in Fig.~\ref{fig:feynman_diag_cs}. The main processes contributing to the indirect detection cross-section and relic density are annihilation processes, where two DM particles annihilate into CP-even Higgs bosons and are shown in the appendix in Fig.~\ref{fig:feynman_diag_relden}.
\subsection*{Comparison with $Z^{\prime}_2$ conserved case}
We now compare the $Z^{\prime}_2$ conserving and $Z^{\prime}_2$ broken symmetric 2HDMS as in Table~\ref{tab:comparison}. In the former case, the singlet scalar does not develop a \textit{vev} and does not mix with the Higgs doublets. Therefore, it  is characterized by the presence of five physical Higgs bosons (as in the 2HDM), i.e, two CP-even Higgses, a CP-odd  Higgs and a pair of charged Higgses in addition to the complex scalar giving rise to the DM candidate.  In the case, where the singlet obtains a \text{vev}, the scalar component of the singlet mixes with the Higgses leading to an enlarged Higgs sector consisting of six physical Higgs bosons, i.e, three CP-even Higgses, one CP-odd Higgs and a pair of charged Higgses in addition to a pseudoscalar DM candidate.  Consequently, there are two extra mixing angles in the scalar sector.
\begin{table}[ht!]
    \centering
    \scriptsize
    \begin{tabular}{|c|c|c|}
        \hline
         & $Z_2'$ breaking ($v_S \neq 0$) & $Z_2'$ conserving ($v_S = 0$) \\
        \hline
        No. of free Parameters & 15 & 15 \\
        DM Candidate & $A_S$ &  $A_S$,$h_S$ \\
        DM Mass & $m_{A_S}^2 =-(2m_S'^2 + \frac{2}{3} \lambda_1'' v_S^2$  & $m^2_{h_s/A_S}=m^2_S\pm m^{2\prime}_S +(\lambda^{\prime}_1\pm2\lambda^{\prime}_4)\frac{v^2_1}{2}$ \\
        & $ \quad \quad \quad \quad + 2(\lambda_4' v_1^2 + \lambda_5'v_2^2))$ & $+(\lambda^{\prime}_2\pm2\lambda^{\prime}_5)\frac{v^2_2}{2}$ \\
        Particle Spectrum & 1 charged Higgs,  & 1 charged Higgs,  \\
        & 1 charged Goldstone,  & 1 charged Goldstone,  \\
        & 3 scalar Higgs,  & 2 scalar Higgs,  \\
        & 1 pseudo scalar Higgs,  & 1 pseudo scalar Higgs,  \\
        & 1 pseudo scalar Goldstone,  & 1 pseudo scalar Goldstone,  \\
        & 1 pseudo scalar DM candidate & 2 DM candidates \\
        Scalar Mixing Angles & $\alpha_{1,2,3}$ & $\alpha$ \\
        \hline
    \end{tabular}
    \caption{Differences between 2HDMS $Z_2'$ breaking and $Z_2'$ conserving case.}
    \label{tab:comparison}
\end{table} 
\section{Constraints}
\label{sec:const}
\subsection{Theoretical Constraints}
\begin{itemize}
\item \textbf{Boundedness-from-Below (bfb) Conditions} 
The bfb conditions essentially demand the positivity of the potential for sufficiently
large values of the field. Since at large field values
the potential is dominated by the quartic terms, this condition puts significant constraints on the quartic couplings of the scalar potential. The required conditions for the 2HDM and its extension with a real singlet scalar have been calculated in~\cite{Klimenko:1984qx,Nie:1998yn,Drozd:2014yla}.  
In this work, we have derived the conditions for the scalar potential pertaining to the complex scalar singlet extension of 2HDM to be bounded from below. 
They can be found by writing the minimum of the part of the potential containing only terms with four orders of fields in matrix form using copositivity conditions~\cite{Kannike:2012pe}.
In that work, the 2HDM potential has been considered, but from the steps described there, the conditions for other potentials can be derived as follows. \\
We start with the potential $V$ from eq.~\ref{eq:cspot} and take only terms containing 4 orders of fields into consideration. This part of the potential is denoted $V_4$.\footnote{The terms containing 2 orders of fields do not have to be considered for the calculation of the bfb conditions, since for the behaviour of the potential at infinity, terms with 2 orders of fields can be neglected compared to terms with 4 orders of fields.} From this, the minimum $min[V_4]$ has to be found and written in matrix form in the basis $X=\begin{pmatrix} \Phi_1^\dagger \Phi_1 , & \Phi_2^\dagger \Phi_2 , & \rho_S^2 , & \eta_S^2 \end{pmatrix}^T$, with $S=\rho_S + i\eta_S$:
\begin{align}
    min[V_4]
    &=X^T \frac{1}{2} \underbrace{\begin{pmatrix}
    \lambda_1 & 
    \lambda_3 + \rho^2(\lambda_4 -|\lambda_5|) & 
    \lambda_1' + 2\lambda_4' & 
    \lambda_1' - 2\lambda_4' \\
    \lambda_3 + \rho^2(\lambda_4 -|\lambda_5|) & 
    \lambda_2 & 
    \lambda_2' + 2\lambda_5' & 
    \lambda_2' - 2\lambda_5' \\
    \lambda_1' + 2\lambda_4' & 
    \lambda_2' + 2\lambda_5' & 
    \frac{5\lambda_1'' + 3\lambda_3''}{6} & 
    \frac{-\lambda_1'' + \lambda_3''}{2} \\
    \lambda_1' - 2\lambda_4' & 
    \lambda_2' - 2\lambda_5' & 
    \frac{-\lambda_1'' + \lambda_3''}{2} & 
    \frac{-\lambda_1'' + \lambda_3''}{2}
    \end{pmatrix}}_{A} X \nonumber \\
    &= \frac{1}{2} X^T A X ,
    \label{bfbeq}
\end{align}
where two cases are distinguished:
\begin{align*}
    \text{case 1: } (\lambda_4 - |\lambda_5|) &\geq 0 \quad \Rightarrow \quad \min[V_4]=V_4|_{\rho=0}\\
    \text{case 2: } (\lambda_4 - |\lambda_5|) &< 0 \quad \Rightarrow \quad \min[V_4]=V_4|_{\rho=1} .
\end{align*}
Requiring the potential to be bounded from below then is equivalent to requiring the matrix $A$ to be copositive. $A$ is a symmetric $4 \times 4$ matrix. In order to derive the copositivity conditions the Cottle-Habetler-Lemke theorem~\cite{Cottle1970OnCO} can be followed, as described in \cite{Kannike:2012pe}. This can be done in two steps:
\begin{itemize}
    \item The order 3 principal submatrices of $A$ are required to be copositive. (The order 3 principal submatrices are obtained by deleting the $i$-th row and column from $A$, $i=1,2,3,4$. This results in 4 symmetric $3 \times 3$ matrices.) \\
    The explicit copositivity conditions for a symmetric order 3 matrix $B$ with entries $b_{ij}$, $i,j=1,2,3$ can be found in~\cite[eq. (5) and (6)]{Kannike:2012pe} and are:
    \begin{align}
        &b_{11} \geq 0 , \quad b_{22} \geq 0 , \quad b_{33} \geq 0,  \label{bfb1}\\ 
        &\Bar{b_{12}} = b_{12} + \sqrt{b_{11}b_{22}} \geq 0,  \\
        &\Bar{b_{13}} = b_{13} + \sqrt{b_{11}b_{33}} \geq 0,  \\
        &\Bar{b_{23}} = b_{23} + \sqrt{b_{22}b_{33}} \geq 0, \\
        &\sqrt{b_{11}b_{22}b_{33}} + b_{12}\sqrt{b_{33}} + b_{13}\sqrt{b_{22}} + b_{23}\sqrt{b_{11}} + \sqrt{2 \Bar{b_{12}} \Bar{b_{13}} \Bar{b_{23}}} \geq 0. \label{bfb5}
    \end{align}
    \item The matrix $A$ has to satisfy: $\det(A) \geq 0 \quad \lor \quad (\text{adj} A)_{ij} < 0$, for some $i,j$. \\
    The adjugate of A is defined as the transpose of the cofactor matrix: $(\text{adj} A)_{ij} = (-1)^{i+j}D_{ji}$, with $D_{ij}$ being the determinant of the submatrix that is obtained by deleting the $i$-th row and $j$-th column from $A$.
\end{itemize}
We implemented the conditions above in python using \texttt{numpy}~\cite{numpy:Harris_2020} to check for each point whether bfb is satisfied or not.
\item \textbf{Vacuum Stability}
The requirement of vacuum stability at the EW
scale places additional constraints on the
parameter space. The strongest constraint comes from demanding the EW vacuum to be the global minimum of the full scalar potential. In this case the EW-vacuum will be absolutely stable. The absolute stability implies that there exist no charge or CP-breaking minima, or non-EW vacuum lower than the EW vacuum. However this constraint can be relaxed if one demands, even if there exist any of those unphysical minima lower than EW vacuum, the transition time for EW vacuum to the unphysical minima is higher than the age of the universe, thus indicating metastability. In that case the EW-vacuum is sufficiently long-lived, albeit not absolutely stable. We consider for a given parameter point a EW vacuum short-lived and a deeper minima potentially dangerous if the quantity called `bounce action'~\cite{Coleman:1977py,Callan:1977pt,Adams:1993zs} $B < 390$. 
In our model, although the bfb puts strong constraints on the quartic couplings, there can be further constraints on the quartic couplings, from the requirement of (meta)stability of the EW vacuum.  Exploration of vacuum stability of the entire model parameter space is beyond the scope of the present work. However, we have ensured that for our chosen analysis benchmark the EW vacuum is absolutely stable.  For our study we have used {\tt EVADE}~\cite{Hollik:2018wrr,Ferreira:2019iqb}, which uses {\tt HOM4SP2}~\cite{Lee2008} to find the tree-level minima and in case of an unphysical deeper minimum, it calculates the bounce action using straight path approximation~\cite{Hollik:2018wrr}.
\item \textbf{Tree-Level Unitarity Conditions}
The tree-level unitarity conditions put a limit on the eigenvalues of the scattering matrices between the scalars and the longitudinal components of the gauge bosons. These conditions have been checked for our model using \texttt{SARAH-SPheno} files in \texttt{SPheno-v4.0.5}~\cite{Porod:2003um}  ensuring that all the accepted points obey the condition that the maximal eigenvalue of the scattering matrix is less than $\frac{1}{2}$ (see Ref.~\cite{Goodsell:2018tti} for more details).
\end{itemize}
\subsection{Experimental Constraints} 
The relevant experimental constraints for our study are: 
\begin{itemize}  
  \item The  second lightest CP-even Higgs, $h_2$ is the SM-like Higgs with mass, $m_{h_2} = 125.25\pm 0.17$ GeV within the experimental error~\cite{ATLAS-CONF-2020-005}.
  \item The invisible decay width of the  SM-like Higgs to the DM candidate $A_S$, is constrained by ATLAS and CMS as below, 
  \begin{align*}
  \begin{split}
  BR(h_{2} \rightarrow A_S A_S) &\leq 0.07^{+0.030}_{-0.022}    \text{  (ATLAS)}\text{~\cite{ATLAS:2023tkt}}
  \\&\leq 0.15 \text{ (CMS)} \text{~\cite{CMS:2023sdw}}.
  \end{split}
  \end{align*}
  \item Flavor physics constraints, namely $BR(b \rightarrow s \gamma$) = (3.55$\pm$0.24$\pm$0.09)$\times 10^{-4}$~\cite{Lees:2012ym}, $BR(B_s \rightarrow \mu^+\mu^-$)=(3.2$^{+1.4\\+0.5}_{-1.2\\-0.3})\times10^{-9}$~\cite{Aaij:2013aka,Chatrchyan:2013bka}.
  The benchmark point is also  within the upper limit of $\Delta(g-2)_{\mu}( = 261 (63) (48) \times 10^{-11})$~\cite{ParticleDataGroup:2020ssz}.
  \item The benchmark point  also satisfies the electroweak precision test constraints on the $STU$ parameters, where $S = 0.02 \pm 0.1$, $T =0.07\pm0.12$, $U = 0.00\pm0.09$~\cite{10.1093/ptep/ptaa104} and the model predictions of $STU$ parameters are obtained from~\cite{Grimus:2007if,Grimus:2008nb}.
  \item The relic density upper limit from \texttt{PLANCK} data, i.e, $\Omega h^2 =  0.1191 \pm 0.0010  $~\cite{Aghanim:2018eyx} is adhered to. 
  \item DM-nucleon spin independent cross sections from \texttt{LUX-ZEPLIN  (LZ)}~\cite{LZ:2022ufs} and indirect detection constraints from \texttt{Fermi-LAT}~\cite{Fermi-LAT:2011vow,Fermi-LAT:2016uux}. Here we would like to mention that, even for the parameter points which lead to underabundance of relic density, we have not rescaled the direct or indirect detection cross-sections with the ratio of actual relic abundance of our postulated DM candidate and the total observed relic abundance of the universe~\cite{Barger:2008jx,Biekotter:2021ovi,Belanger:2022esk}. In that way, our study is quite conservative and an even more relaxed parameter space can open up with the aforementioned rescaling.
  \item The constraints from LEP~\cite{Abbiendi:2013hk} and  ATLAS$/$CMS searches on the heavy Higgs searches~\cite{higgssumatlas,higgssumcms} and the 125 GeV Higgs signal strength measurements~\cite{ATLAS-CONF-2020-027} are taken into account.
\end{itemize}
The model files  are generated using \texttt{SARAH-v4.14.3}~\cite{Staub:2013tta} and the particle spectra and decays are generated using \texttt{SPheno-v4.0.5}~\cite{Porod:2003um}.\footnote{The model files and benchmarks associated with this paper are available at Ref.~\cite{dutta_2024_10569080}.} The DM observables have been computed using~\texttt{micrOmegas-v5.2.13}~\cite{mco2} and the Higgs constraints are checked using \texttt{HiggsTools}~\cite{Bahl:2022igd,Bechtle:2013xfa,Bechtle:2013wla,Bechtle:2020pkv,Bechtle:2020uwn}. \\
The constraints are applied as binary-cut. Hence only parameter points which are allowed by all constraints are considered allowed.
\section{Dark Matter Phenomenology}
\label{sec:dm} 
In this section, we discuss the impact of DM observables on the parameter space of 2HDMS.
We start with briefly discussing the implications of the 95 GeV excess observed at CMS and LEP on the 2HDMS parameters. The observed signal strengths of 95 GeV excess for LEP in the $b\bar{b}$ mode ($\sim 2 \sigma$)~\cite{LEPWorkingGroupforHiggsbosonsearches:2003ing} and LHC in the $\gamma\gamma$ mode ($\sim 3 \sigma$)~\cite{CMS:2023yay,ATLAS:2023jzc} are,
 \begin{equation}
 \mu^{b\bar{b}}_\mathrm{LEP} =   0.117^{+0.057}_{-0.057}, \qquad
 \mu^{\gamma\gamma}_\mathrm{LHC-combined} =    0.24^{+0.09}_{-0.08},
\end{equation} 
where the ATLAS and CMS results are combined following Ref.~\cite{Biekotter:2023oen}.

From Ref.~\cite{Heinemeyer:2021msz}, in the Type II 2HDMS the reduced couplings  of singlet-like Higgs
$h_1$ as follows,
\begin{eqnarray}
c_{h_1t\bar{t}} = \frac{R_{12}}{\sin \beta}, \\
c_{h_1b\bar{b}} = \frac{R_{11}}{\cos \beta}, \\
c_{h_1\tau\tau} = \frac{R_{11}}{\cos \beta}, \\
c_{h_1 VV} = \cos \beta R_{11} + \sin \beta R_{12},
\end{eqnarray}
where $R_{ij}$ refers to the elements of the rotation matrix. According to Ref.~\cite{Heinemeyer:2021msz}, the coupling $c_{h_1b\bar{b}}$ can not strongly affect the $h_1\rightarrow b\bar{b}$ branching ratio, while the total width of $h_1$ is dominated by the $c_{h_1b\bar{b}}$ coupling. In this case, the Higgs strahlung production plays the most important role of LEP signal strength, and the $h_1\rightarrow\gamma\gamma$ branching ratio is strongly dependent on the $h_1$ total width. Therefore, the signal strengths of the observed 95 GeV excess may be correlated with the reduced couplings and mixing angles as below:
\begin{equation}
    \mu^\mathrm{2HDMS}_{b\bar{b}} \propto |c_{h_1 VV}|^2,
\end{equation}
and 
\begin{equation}
   \mu^\mathrm{2HDMS}_{\gamma\gamma} \propto \frac{(|c_{h_1t\bar{t}}|)^2}{(|c_{h_1b\bar{b}}|)^2}  
\propto (\frac{\tan \alpha_1}{\tan \beta})^2,
\end{equation}
where $\mu^\mathrm{2HDMS}_{b\bar{b}}$ and $\mu^\mathrm{2HDMS}_{\gamma\gamma}$ are the  signal strengths of the $b\bar{b}$ and $\gamma\gamma$ channels computed in 2HDMS. 
In order to obtain  a benchmark point consistent with  all theoretical constraints including unitarity and bfb, and experimental constraints from DM, Higgs and collider constraints on the heavy Higgs bosons and  the observed 95 GeV excess, we set up a global scan keeping the Higgs sector fixed such that the lightest CP-even Higgs, $h_1$ is the 95 GeV Higgs while $h_2$ is the SM-like 125 GeV Higgs. 
The parameters are scanned , using random sampling,  over the following ranges, 
\begin{center}
 $\tan \beta=10$, $\frac{\tan\beta}{\tan\alpha_1}=0.35$, $\alpha_2=-1.2$, $\beta-\alpha_1-\alpha_3=-[1.54,1.6]$, \bigskip  
$m_{h_1}=95$ GeV, $m_{h_2}=125$ GeV, $m_{H^{\pm}}=m_A=m_{h_3}=900$ GeV, $v_s = [100,1000]$ GeV, \quad  \\
 $m_{A_s} = [48,800]$ GeV, $m^{2\prime}_S=[0,10^6]$ GeV$^2$, $\lambda^{\prime}_4= [-3:3]$, $\lambda^{\prime}_5=[-3:3]$.
\end{center} 
Subsequently, we choose a benchmark point \textbf{BP1} as shown in Table~\ref{tab:bp1} satisfying all the theoretical and experimental constraints and expressed in the chosen mass basis parameters as shown in eq.~\ref{eq:freepar_mass}. While scanning the parameter space, we varied $\lambda^{\prime}_4$ and $\lambda^{\prime}_5$. However, for the chosen benchmark, we have re-expressed them in terms of $\delta_{14}'$ and $\delta_{25}'$, since they are the relevant parameters in question. We chose a rather high $\tan\beta$ value in order to obey the DM direct detection cross section bounds. The dependence of direct detection cross-section on $\tan\beta$ enters via DM-scalar coupling (Eq.~\ref{eq:DM_couplings_tril} \ref{eq:DM_couplings_15dof_mass_basis}) as well as the relevant Yukawa couplings. In this case, due to the presence of multiple scalars leading to $t$-channel mediation and the possible interference effects between them, it is difficult to isolate the $\tan\beta$-behavior analytically. However, in the following section we present the $\tan\beta$-dependence in corresponding numerical scans (Please see Fig.~\ref{fig:infl_vS_tanbeta} in particular.). Furthermore, the low $m_{A}$ region would be excluded by the LHC $A\rightarrow\tau^+\tau^-$ searches~\cite{CMS:2018rmh,ATLAS:2020zms,CMS:2022rbd} for high $\tan\beta$ values. Therefore, we chose $m_{A}$, $m_{h_3}$ and $m_{H^\pm}$ appropriately heavy.
\begin{table}[h!]
    \centering
    \addtolength{\tabcolsep}{-3pt}
    \small
    \begin{tabular}{|c|c|c|c|c|}
        \hline
        $m_{h_1}$ & $m_{h_2}$ & $m_{h_3}$ & $m_A$ & $m_{A_S}$ \\ 
        \hline
        $95 \, \text{GeV}$ & $125.09 \, \text{GeV}$ & $900 \, \text{GeV}$ & $900 \, \text{GeV}$ & $325.86 \, \text{GeV}$ \\
        \hline
        $m_{H^\pm}$ & $m_{S}'^2$ & $\delta_{14}'$ & $\delta_{25}'$ &  $\tan(\beta)$\\
        \hline
        $900 \, \text{GeV}$ & $ -4.809 \times 10^4 \, \text{GeV}^2$ & $-9.6958$ & $0.2475$ &  $10$\\ 
        \hline  
        $v_S$ & $c_{h_1 bb}$ & $c_{h_1 tt}$ & $\mathrm{alignm}$ & $\Tilde{\mu}^2$  \\
        \hline
        $ 239.86 \, \text{GeV}$ &  0.2096  & 0.4192  & $0.9998$ & $ 8.128 \times 10^5 \, \text{GeV}^2$  \\
        \hline 
    \end{tabular}
    \caption{The benchmark point \textbf{BP1} in the mass basis.}
    \label{tab:bp1}
\end{table} \\
In the following sections we scan around \textbf{BP1}, using fixed grids and varying two parameters at the same time. We use about 100 steps for each parameter. This leaves a total of $100 \times 100 = 10000 $ points for each scan.
\subsection{Impact of 95 GeV Excess}
\begin{figure}[ht]
     \centering
     \includegraphics[scale=0.45]{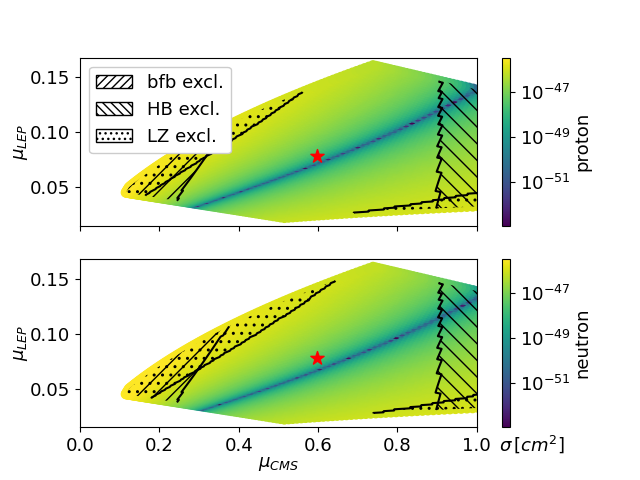}
     \includegraphics[scale=0.45]{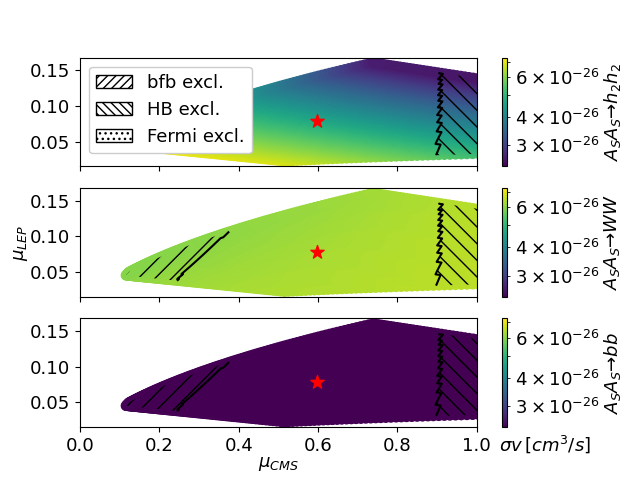}\\
     \includegraphics[scale=0.45]{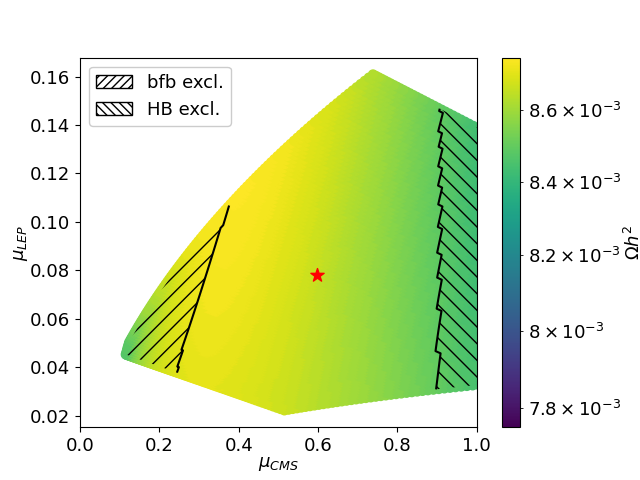}
     \includegraphics[scale=0.45]{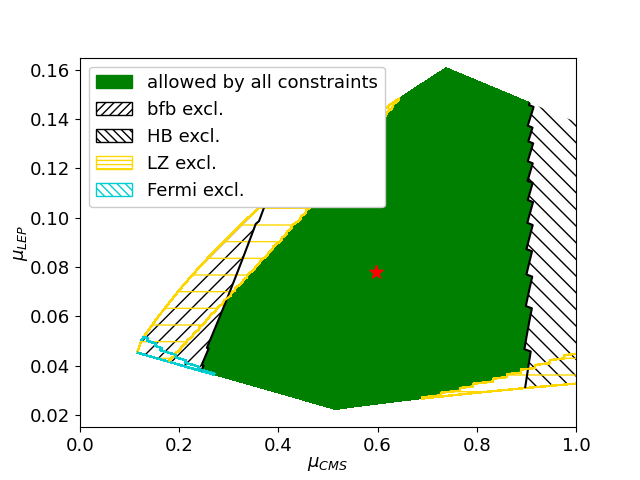}\\
     \caption{Variation of $\mu_{\text{CMS}}-\mu_{\text{LEP}}$ to fit the 95 GeV excess signal strength for $h_1$. The relevant constraints which stringently restrict the allowed regions are shown in the plots. The coloured palette on the z-axis denote the spin-independent direct detection DM-proton/neutron cross-section (top left), indirect detection DM annihilation cross-section (top right), relic density (bottom left) and the allowed parameter regions under combining all constraints (bottom right). \textbf{BP1},  in  Table~\ref{tab:bp1}, is marked with a red star.} 
     \label{fig:muLEPmuCMS}
\end{figure}

The strength of the 95 GeV excess in the $\gamma\gamma$ channel at CMS ($\sim$ 2.9 $\sigma$) at LHC and in the $b\bar{b}$ mode at LEP ($\sim$ 2$\sigma$) can be fit in  Type II 2HDMS  and has been  previously studied for different symmetries~\cite{Heinemeyer:2021msz,Biekotter:2021ovi,Biekotter:2023oen}. Recently a slight excess of $\sim 1.7 \sigma$ has also been observed in the $\gamma\gamma$ mode at ATLAS~\cite{ATLAS-CONF-2023-035}. It has been studied in the context of 2HDMS in Ref.~\cite{Biekotter:2023oen}. In this work, we focus mainly on the excess observed at CMS in the $\gamma\gamma$ channel and at LEP in the $b\bar{b}$ mode. \\
Fig.~\ref{fig:muLEPmuCMS} shows the allowed $\mu_\mathrm{CMS}-\mu_\mathrm{LEP}$ plane subject to the theoretical constraints from bfb and tree-level unitarity and experimental constraints from the Higgs sector as well as constraints from DM observables, namely, spin-independent direct detection cross-section for scattering on protons and on neutrons, indirect detection cross-section for DM annihilation (for the channels $h_2 h_2$, $W W$ and $b \Bar{b}$) and relic density. The data was obtained by varying the reduced couplings $c_{h_1 bb}$ and $c_{h_1 tt}$ as in Table~\ref{tab:ch1bb_ch1tt} and keeping all other parameters fixed to \textbf{BP1}\footnote{The range for $c_{h_1 bb}$ and $c_{h_1 tt}$ was chosen such that the constraints from Ref.~\cite{Heinemeyer:2021msz} are respected. This results in some areas in the $\mu_\mathrm{CMS}-\mu_\mathrm{LEP}$ plane being   white, as those values are not allowed by the constraints from the observed 95 GeV excess.}. 
\begin{table}[ht]
      \centering
      \begin{tabular}{|c|c|}
     \hline
     Parameters & Range\\
     \hline
    $c_{h_1 b b}$ & $[0.0996, 0.320]$ \\
    $c_{h_1 t t}$ & $[0.309, 0.529]$ \\
    \hline
     \end{tabular}
     \caption{List of parameters varied. The rest of the parameters are kept fixed to \textbf{BP1}, Table~\ref{tab:bp1}, as discussed in the text.}
     \label{tab:ch1bb_ch1tt}
\end{table} 
The signal strengths $\mu_\mathrm{CMS}$ and $\mu_\mathrm{LEP}$ were calculated for the different values of $c_{h_1 bb}$ and $c_{h_1 tt}$ and are shown on the x-axis and y-axis, respectively.
As can be seen in Fig.~\ref{fig:muLEPmuCMS} from the top left plot the direct detection cross-section has a minimum along an arched line. Close to this line lies \textbf{BP1}, marked with a red star. The regions where the cross-section grows too large and exceeds the upper bounds from LUX-ZEPLIN are excluded and shown as a dotted area. This occurs due to insufficient cancellations between the different contributions to the spin independent direct detection cross-section from $h_1, h_2$ and $h_3$.  \\
From the top right plot one can see how the indirect detection cross-sections for the channels $h_2 h_2$, $W W$ and $b \Bar{b}$ behave. The cross-section for the $h_2 h_2$ channel decreases with $\mu_\mathrm{LEP}$, whereas the other two channels do not show large changes. Due to the $h_2 h_2$ channel dominating for the chosen benchmark \textbf{BP1} the variations of the other channels are on a smaller scale. 
Only a thin line in the $h_2 h_2$ channel with low $\mu_\mathrm{CMS}$ and low $\mu_\mathrm{LEP}$ is excluded by Fermi-LAT data. \\
In the bottom left plot the relic density is shown. It  shows small changes and reaches the highest values around $\mu_\mathrm{CMS} \approx 0.3$. However, it is very low across the whole region and always remains underabundant. \\
In the bottom right plot all the constraints are combined, revealing the allowed region, shown in green.

\subsection{Impact of free Parameters on Dark Matter Observables}
\label{subsec:dm13}

Before we discuss the impact of the free parameters of our model on DM observables in detail, a few important comments should be made. The DM phenomenology in our model is crucially coupled with the chosen symmetries of the DM sector, which is a discrete $Z_2'$ symmetry in our case. A relevant comparison can be made with several earlier works~\cite{Gross:2017dan,Biekotter:2021ovi,Biekotter:2022bxp} with a complex singlet, where the DM sector is stabilized by the imposition of $U(1)$ symmetry (softly broken) instead. In those models, the spontaneous breaking of the continuous $U(1)$ symmetry by the singlet {\it vev} gives rise to a pseudo Nambu-Goldstone Boson (pNGB) DM, which can evade direct detection constraints to a large extent and these models are difficult to probe even in future direct search experiments. We would reiterate that this is not the case for us. Spontaneous breaking of $Z_2'$ symmetry does not lead to pNGB DM, due to the presence of $U(1)$ breaking terms allowed by $Z_2'$ symmetry. Therefore in our case the direct detection bounds constrain our parameter space significantly, as we will see below. Furthermore, we will have parts of parameter space in our models that will necessarily come under the scanner of future direct search experiments.
Now we study the impact of the free parameters on the DM observable  considering the benchmark point \textbf{BP1}.
\subsubsection{Influence of $\delta_{14}'$ and $\delta_{25}'$}
\label{subsec:l4pl5p}
\begin{figure}
    \centering
    \includegraphics[scale=0.45]{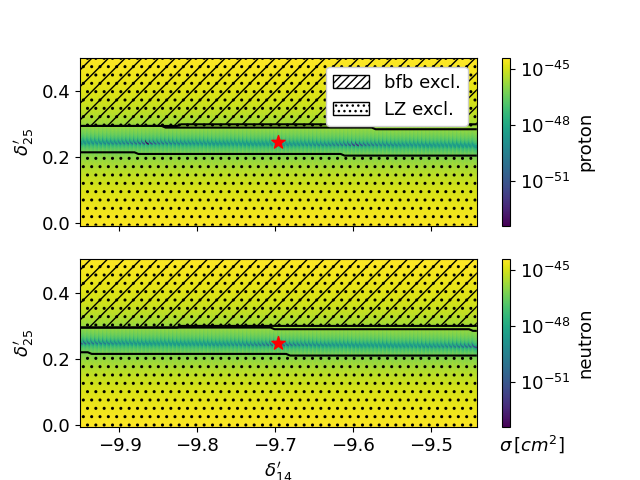}
    \includegraphics[scale=0.45]{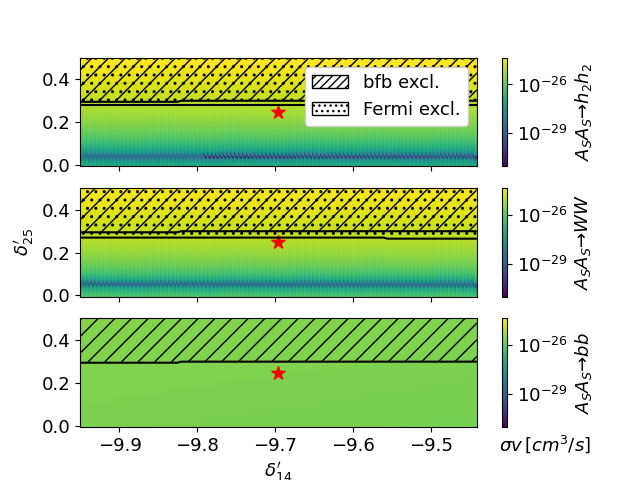}\\
    \includegraphics[scale=0.45]{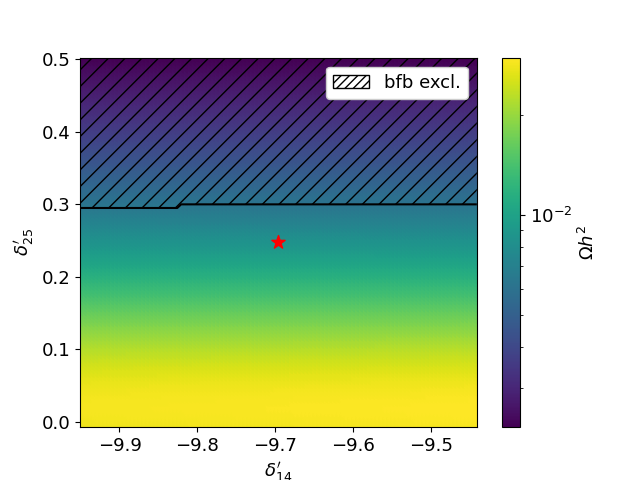}
    \includegraphics[scale=0.45]{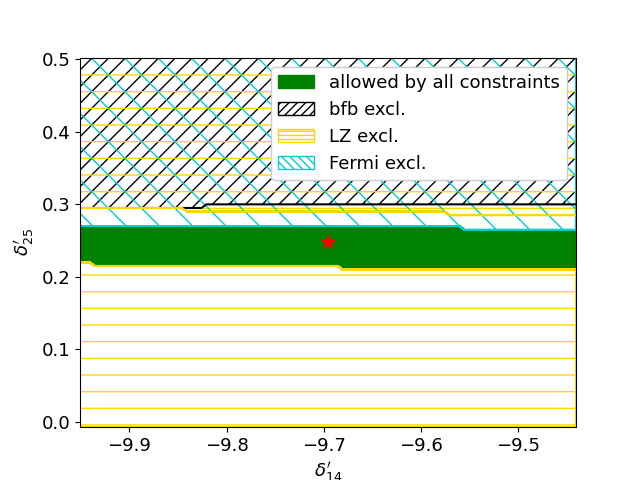}
    \caption{Influence of $\delta_{14}'=\lambda_4' - \lambda_1'$ and $\delta_{25}'=\lambda_2' - \lambda_5'$. The coloured palette on the z-axis denote the spin-independent direct detection DM-proton/neutron cross-section (top left), indirect detection DM annihilation cross-section (top right), relic density (bottom left) and the allowed parameter regions under combining all constraints (bottom right). \textbf{BP1}, in  Table~\ref{tab:bp1},  is marked with a red star.}
    \label{fig:infl_dl14p_dl25p}
\end{figure}

 We defined earlier two variables $\delta^{\prime}_{14} $ and $\delta^{\prime}_{25}$ where,
\begin{align}
    \delta^{\prime}_{14}=\lambda^{\prime}_{4}-\lambda^{\prime}_{1}, \\
    \delta^{\prime}_{25}=\lambda^{\prime}_{5}-\lambda^{\prime}_{2}.
\end{align} 
Starting with the parameters of \textbf{BP1} we vary $\delta^{\prime}_{14}$ and $\delta^{\prime}_{25}$ as in Table~\ref{tab:dl14p_dl25}.
\begin{table}[ht]
      \centering
      \begin{tabular}{|c|c|}
     \hline
     Parameters & Range\\
     \hline
    $\delta^{\prime}_{25}$ & $[-0.01, 0.50]$ \\
    $\delta^{\prime}_{14}$ & $[-9.95, -9.44]$ \\
    \hline
     \end{tabular}
     \caption{List of parameters varied. The rest of the parameters are kept fixed to \textbf{BP1} as discussed in the text.}
     \label{tab:dl14p_dl25}
\end{table}
The results are shown in Fig.~\ref{fig:infl_dl14p_dl25p}, where $\delta^{\prime}_{14}$ is varied along the x-axis and $\delta^{\prime}_{25}$ is varied along the y-axis. The influence on the direct detection cross section for scattering on protons and on neutrons, indirect detection cross section for DM annihilation and relic density are shown in the coloured palette. \textbf{BP1} is marked with a red star and the excluded regions are shown as hatched areas. The plot on the bottom right shows a summary of all exclusions and the allowed region in green.
As can be seen $\delta^{\prime}_{14}$ does not have a strong impact on the observables in this benchmark. This can be explained by the choice of $\tan\beta$. From eq.~\ref{eq:DM_couplings_15dof_mass_basis} it can be seen that by setting $\tan\beta=10$ (see Table~\ref{tab:bp1})   $\cos\beta$ suppresses the influence of $\delta^{\prime}_{14}$ and  $\sin\beta$ enhances the influence of $\delta^{\prime}_{25}$. For the direct detection cross-section we see a dip around $\delta^{\prime}_{25} \approx 0.25$. Such a dip can in principle occur due to cancellation between the elastic scattering amplitudes involving $h_1$ and $h_2$ (owing to their closeness in mass), if their contribution comes with opposite signs. Such cancellations also arise in simple Higgs portal models such as the complex scalar extended SM~\cite{Gross:2017dan}. We have seen that for our chosen parameter space, they indeed come with opposite signs. The areas away from the dip are above the upper bounds from LUX-ZEPLIN and are therefore excluded, shown as dotted regions.  \\
For the indirect detection cross-section in the $h_2 h_2$ channel and the $W W$ channel one can see dips around $\delta^{\prime}_{25}\approx 0.05$. For higher values of $\delta^{\prime}_{25}$ the cross-sections grow. 
The areas where they get too high,i.e. $\delta^{\prime}_{25} \approx 0.3$, are excluded by constraints from Fermi-LAT, shown again as dotted regions. The $b \Bar{b}$ channel does not show large changes. \\
The relic density, on the other hand, falls with increasing $\delta^{\prime}_{25}$ as shown in the bottom left plot remaining below the upper bound from PLANCK.  This can be explained as follows, as can be seen from eq.~\ref{eq:DM_couplings_15dof_mass_basis} and \ref{eq:DM_couplings_15dof_mass_basis_quatr}, increasing $\delta_{25}'$ increases the trilinear and the quatrilinear DM couplings. This causes more interactions and hence more annihilation of DM particles. After annihilation there is less DM left in the universe. The relic density drops. \\ 
For all three observables, regions above $\delta^{\prime}_{25}\approx0.3$ are excluded by bfb constraints, shown as hatched regions. There are no regions excluded by unitarity constraints or HiggsBounds, here. Combining all constraints from bfb, unitarity, Higgs and DM, this results in a narrow allowed band between $\delta^{\prime}_{25}\approx 0.2$ and $\delta^{\prime}_{25} \approx 0.3$, shown in green (bottom right plot) in Fig.~\ref{fig:infl_dl14p_dl25p}.
\subsubsection{Influence of $v_S$ and $\tan\beta$}
\begin{figure}[ht]
    \centering
    \includegraphics[scale=0.45]{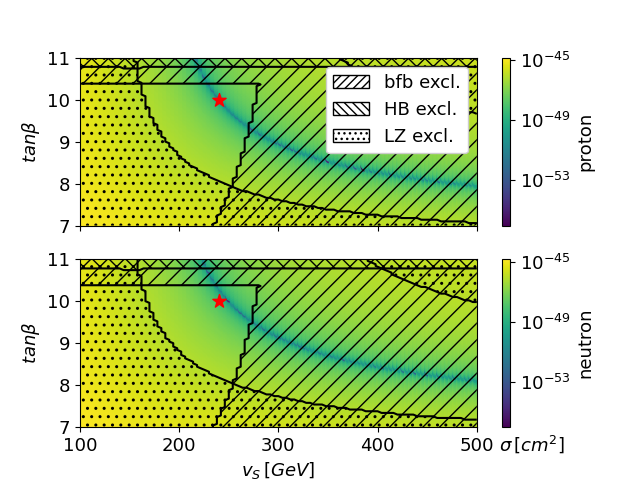}
    \includegraphics[scale=0.45]{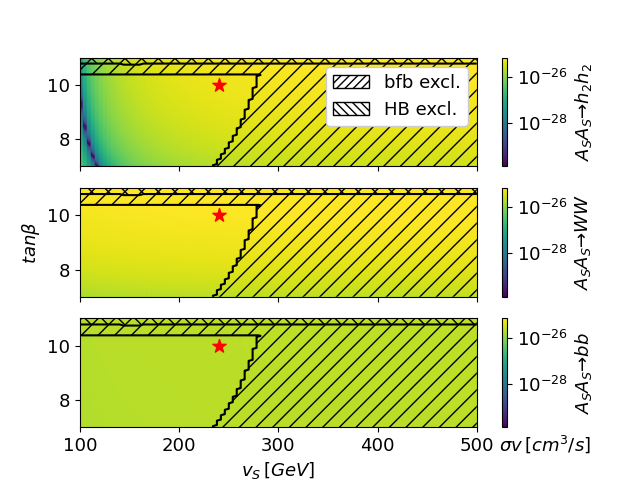}\\
    \includegraphics[scale=0.45]{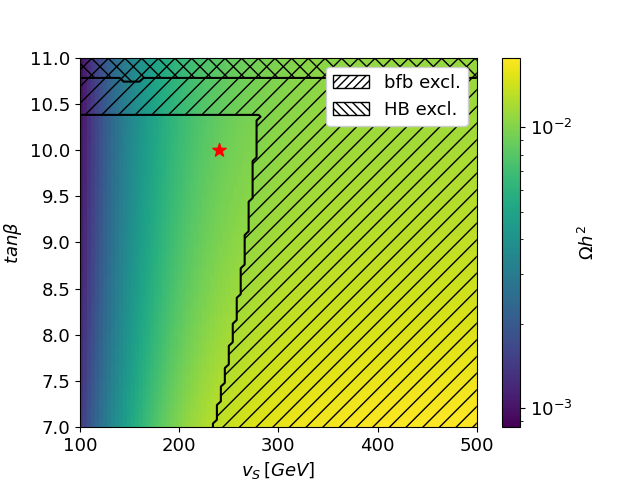}
    \includegraphics[scale=0.45]{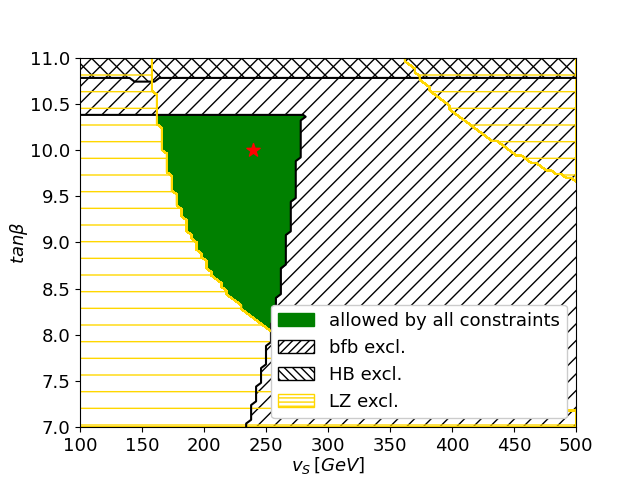}
    \caption{Influence of $v_S$ and $\tan\beta$. The coloured palette on the z-axis denote the spin-independent direct detection DM-proton/neutron cross-section (top left), indirect detection DM annihilation cross-section (top right), relic density (bottom left) and the allowed parameter regions under combining all constraints (bottom right). \textbf{BP1},  in  Table~\ref{tab:bp1},  is marked with a red star.}
    \label{fig:infl_vS_tanbeta}
\end{figure}

In this section, $v_S$ and $\tan\beta$ are varied in the range shown in Table~\ref{tab:vS_tanbeta}. Again the other parameters are fixed to \textbf{BP1}.
\begin{table}[ht]
      \centering
      \begin{tabular}{|c|c|}
     \hline
     Parameters & Range\\
     \hline
    $v_S$ & $[100, 500] \, \text{GeV}$  \\
    $\tan\beta$ & $[9, 11]$ \\
    \hline
     \end{tabular}
     \caption{List of parameters varied. The rest of the parameters are kept fixed to \textbf{BP1} as discussed in the text.}
     \label{tab:vS_tanbeta}
\end{table}
The results can be seen in Fig.~\ref{fig:infl_vS_tanbeta}, where $v_S$ is varied along the x-axis and $\tan\beta$ is varied along the y-axis. The influence on the DM observables, as well as excluded and allowed regions are shown in the same manner as in the previous plots.
As can be seen, both $v_S$ and $\tan\beta$ have an impact on the direct detection cross-section, which results in a dip visible as an arched line around \textbf{BP1}. The direct detection cross-section is a combined contribution from $h_1$(95 GeV) and $h_2$(125 GeV) mediated diagrams in particular. From  Eq.~\ref{eq:DM_couplings_tril} and \ref{eq:DM_couplings_15dof_mass_basis} one can see the dependence of DM-portal coupling to the aforesaid scalars and the direct detection cross-section hits a minimum(the dip) along the arched line, very close to which lies our \textbf{BP1}. Areas away from this dip have a higher cross-section and are excluded by LUX-ZEPLIN, shown as dotted regions. \\
However $\tan\beta$ does not seem to have a large impact on the indirect detection cross-section. The influence of $v_S$, increasing the cross-section, can be seen in the $h_2 h_2$ channel. The other two channels do not show large changes. There are no areas excluded by Fermi-LAT here.  \\
For the relic density again $\tan\beta$ does not seem to have a large impact. However $v_S$ has an impact, increasing the relic density. This can be understood by looking at eq.~\ref{eq:DM_couplings_15dof_mass_basis} and \ref{eq:DM_couplings_15dof_mass_basis_quatr}, where one can see that $v_S$ appears in the denominator of the trilinear and quatrilinear DM couplings. Hence increasing $v_S$ decreases the couplings, which causes less interaction and less annihilation of DM. There will be more DM left in the universe, which increases the relic density. The influence of $\tan\beta$ is not as visible which can be explained by the fact that $\tan\beta$ was varied only over a small range and the $\sin\beta$ and $\cos\beta$ in eq.~\ref{eq:DM_couplings_15dof_mass_basis} and \ref{eq:DM_couplings_15dof_mass_basis_quatr} do not vary over large ranges as $v_S$. Furthermore with growing $\tan\beta$ the $\sin\beta$ increases, while $\cos\beta$ decreases, hence counterbalancing the effect. There are no areas excluded by PLANCK here. However with an increase in $v_S$ the relic density increases but remains underabundant throughout the range of the scan. \\
For all three observables, regions above $\tan\beta \approx 10.75$ are excluded by HiggsBounds, shown as left directed hatches. Regions above $\tan\beta \approx 10.35$ and above $v_S \approx 260 \, \text{GeV}$ are excluded by bfb constraints, shown as right directed hatches. Combining all constraints results in an allowed region between $\tan\beta \approx 9$ and $\tan\beta \approx 10.35$ and between $v_S \approx 170 \, \text{GeV}$ and $v_S \approx 260 \, \text{GeV}$ as shown in the bottom right plot.
\subsubsection{Influence of $m_{A_S}$ and $m_{S}'^2$}
\begin{figure}[ht]
    \centering
    \includegraphics[scale=0.45]{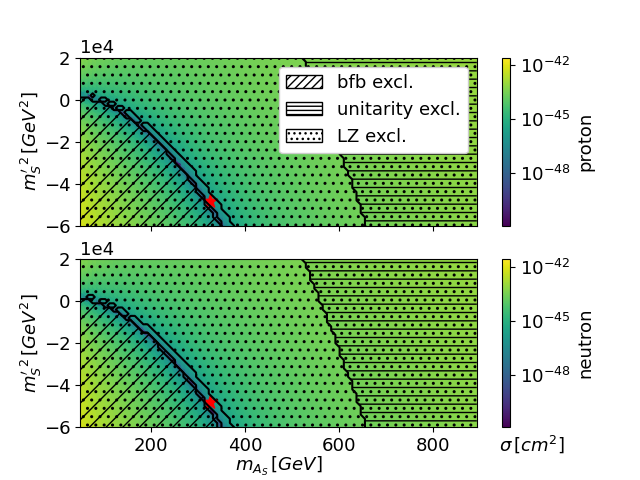}
    \includegraphics[scale=0.45]{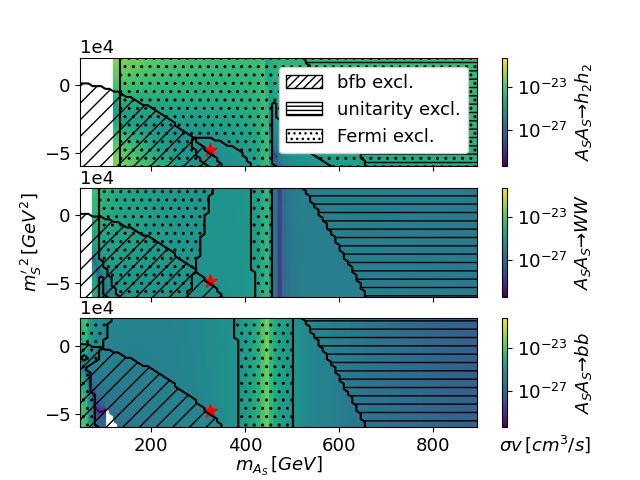}\\
    \includegraphics[scale=0.45]{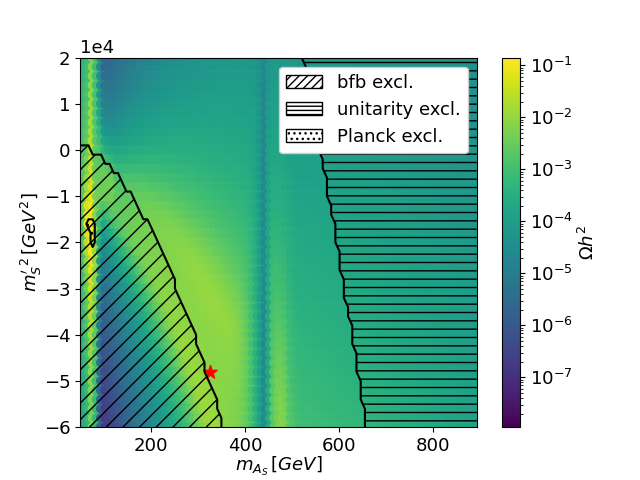}
    \includegraphics[scale=0.45]{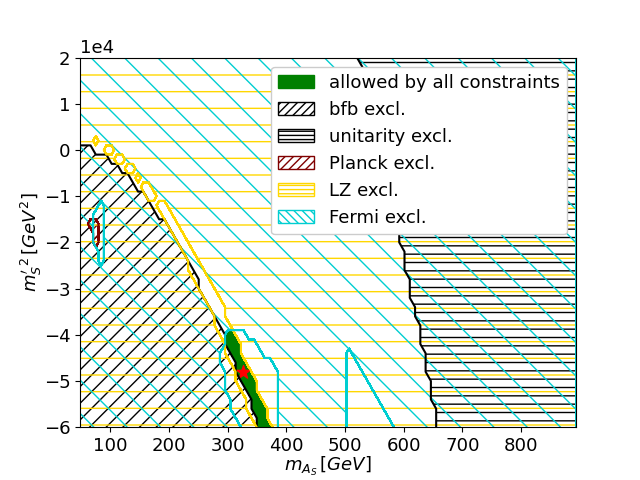}
    \caption{Influence of $m_{A_S}$ and $m_{S}'^2$. The coloured palette on the z-axis denote the spin-independent direct detection DM-proton/neutron cross-section (top left), indirect detection DM annihilation cross-section (top right), relic density (bottom left) and the allowed parameter regions under combining all constraints (bottom right). \textbf{BP1}, in Table~\ref{tab:bp1},  is marked with a red star.}
    \label{fig:infl_mAS_mSp2}
\end{figure}

Finally $m_{A_S}$ and $m_{S}'^2$ are varied in the range shown in Table~\ref{tab:mAS_mSp2}, while the other parameters are kept fixed according to \textbf{BP1}.
\begin{table}[ht]
      \centering
      \begin{tabular}{|c|c|}
     \hline
     Parameters & Range\\
     \hline
    $m_{A_S}$ & $[48, 900] \, \text{GeV}$  \\
    $m_{S}'^2$ & $[-6 \times 10^4, 2 \times 10^4] \, \text{GeV}^2$ \\
    \hline
     \end{tabular}
     \caption{List of parameters varied. The rest of the parameters are kept fixed to \textbf{BP1} as discussed in the text.}
     \label{tab:mAS_mSp2}
\end{table}
The results can be seen in Fig.~\ref{fig:infl_mAS_mSp2}, where $m_{A_S}$ is varied along the x-axis and $m_{S}'^2$ is varied along the y-axis. Again the DM observables, as well as excluded and allowed regions are shown as explained for the previous plots.
As can be seen $m_{A_S}$ and $m_{S}'^2$ change the direct detection cross-section in such a way that a dip appears as an arched line along which \textbf{BP1} lies (similarly as for varying $v_S$ and $\tan\beta$). The mass relation in Eq.~\ref{masseq}, indicates that the arch in the $m_{A_S}$-$m_{S}'^2$ plane would imply a relation between the DM portal couplings, which leads to the minimum in the direct detection cross-section. The areas excluded by LUX-ZEPLIN are again shown as dotted areas, which is almost the whole scanned plane except that thin arched dip line.  \\
The behaviour of the indirect detection cross-section is quite interesting, as one can see a peak in the $b \Bar{b}$ channel around $m_{A_S}\approx 450 \, \text{GeV}$. This is half the mass of the heavy scalars $m_{h_3}=m_{A}=m_{H^\pm}=900 \, \text{GeV}$. An explanation for this peak could be due to the resonant annihilation of two DM particles into the heavy scalar $h_3$, which then, in turn, decays into $b \Bar{b}$. In the $W W $ channel, on the other hand, a strong dip can be observed around $m_{A_S} \approx 475 \, \text{GeV}$   due to the proliferation of $b\bar{b}$, $h_1h_1$, $h_1h_2$ and $t\bar{t}$  channels. 
In all three plots, some white areas can be seen. This is due to \texttt{micrOMEGAs} not returning values in these regions. This can happen when the cross-section is too small. For example in the $h_2 h_2$ channel and the $WW$ channel, where $m_{A_S}<m_{h_2}$ and $m_{A_S}<m_{W}$, the respective annihilation processes are kinematically forbidden. Other channels dominate in this case. The regions excluded by Fermi-LAT are again shown as dotted areas. \\
The influence on the relic density is also  interesting, as maxima can be seen around $m_{A_S}\approx 75 \, \text{GeV}$ and roughly at the arched region, where the direct detection cross-section had its minimum. For lower masses $m_{A_S}\approx 62.5 \, \text{GeV} \approx \frac{m_{h_2}}{2}$ resonant annihilation of two DM particles into one SM Higgs is possible and causes a drop in the relic density. For higher masses $m_{A_S} \gtrapprox 95 \, \text{GeV} \approx m_{h_1}$ the annihilation channel into one light scalar Higgs $h_1$ opens up and also causes the relic density to drop. This could explain why in between those areas the relic density appears higher and looks like a peak. Another interesting feature is the dip at $m_{A_S}\approx 450\, \text{GeV}$, which is half the mass of the heavy scalars $m_{h_3}=m_{A}=m_{H^\pm}=900 \, \text{GeV}$. Here resonant annihilation into those scalars causes the relic density to drop. Almost the whole space is allowed by upper bounds from PLANCK, except a small region in the peak around $m_{A_S}\approx 75 \, \text{GeV}$ and $m_{S}'^2 \approx -18000 \, \text{GeV}^2$, shown as a dotted area. \\
For all three observables, a region with low $m_{A_S}$ and low $m_{S}'^2$ is excluded by bfb constraints, shown as right-directed hatches. Regions above $m_{A_S}\approx 650 \, \text{GeV}$ are excluded by unitarity constraints, shown as horizontal lines. Combining all constraints shows a thin allowed line around $m_{A_S} \approx 350 \, \text{GeV}$ and between $m_{S}'^2 \approx -40000 \, \text{GeV}^2$ and $m_{S}'^2 \approx -60000 \, \text{GeV}^2$ as shown in the bottom right plot. \\
 
\section{Collider Phenomenology}
\label{sec:collider}
\begin{figure}[ht]
    \centering
     \includegraphics[scale=0.5]{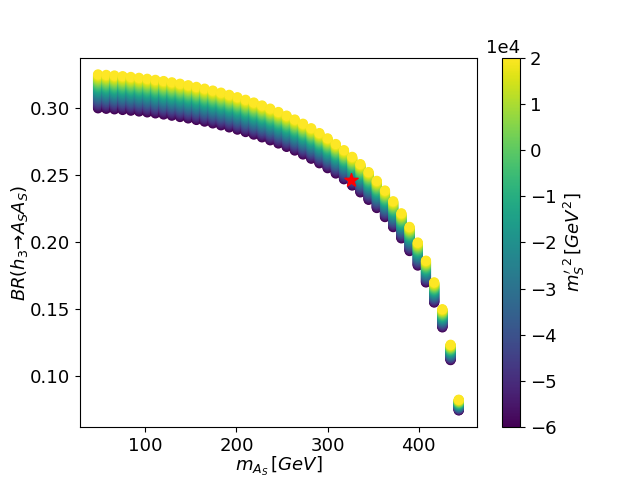}
    \caption{Variation of the invisible branching fraction $BR(h_3 \rightarrow A_S A_S)$ vs. the DM mass $m_{A_S}$ as a function of $m^{2\prime}_S$. The scan parameters for the plot are shown in Table~\ref{tab:mAS_mSp2} while the rest of the parameters are fixed to \textbf{BP1} as discussed in the text. In this plot, only the experimental constraints from the Higgs sector have been considered. }
    \label{fig:brinv}
\end{figure}

In this section,  we discuss the  phenomenology of  2HDMS at present and future colliders. 
Fig.~\ref{fig:brinv}  shows the variation of the invisible branching ratio of the heaviest Higgs into a pair of DM candidates, i.e, $BR(h_3 \rightarrow A_S A_S)$ including constraints from the Higgs sector only in order to understand the dependence of the invisible branching ration on the mass of the DM. Note that the branching ratio decreases with increasing DM mass. This is due to the reduced   phase space for the smaller $h_3-A_S$ mass gap. 
The benchmark \textbf{BP1}  is denoted by a red star in Fig.~\ref{fig:brinv}.  As shown in Table~\ref{tab:decay}, the heavy Higgs $h_3$ dominantly decays  to the $b\bar{b}$ mode followed by the invisible decay to $A_SA_S$ with a  branching fraction of $\simeq 0.25$. It also decays subdominantly into $t\bar{t}, \tau \tau$ and $h_ih_j$ (where i=1,2).  For our purposes, we focus on the detection probability of the invisible mode  and  use \textbf{BP1}  to study the possible signals at LHC and future lepton colliders. 
\begin{table}[ht]
\centering
\begin{tabular}{|c|c|c|c}
\hline
Decay Modes & Branching Ratio (BR)\\
 \hline
$h_3 \rightarrow b \bar{b}$ &0.412 \\
$h_3 \rightarrow A_SA_S$ & 0.247 \\
$h_3 \rightarrow t \bar{t}$ &0.106 \\
$h_3 \rightarrow \tau \tau $& 0.064 \\
$h_3 \rightarrow h_2 h_2$ & 0.061\\
$h_3 \rightarrow h_1 h_2$ & 0.035\\
$h_3 \rightarrow h_1 h_1$ & 0.022\\
\hline
\end{tabular}
\caption{List of the decay modes and the branching ratio for $h_3$ in the benchmark \textbf{BP1}.}
\label{tab:decay}
\end{table}
\subsection*{Simulation set-up}
We generate the parton-level events at $\sqrt{s}=14$ TeV and use \texttt{MG5$\_$aMC$\_$v3.4.1},~\cite{Alwall:2014hca,Alwall:2011uj} followed by showering and hadronization using \texttt{Pythia$\_$v8.3.06}~\cite{Pythia8}. We have used the default parton distribution function \texttt{NNPDF2.3}\cite{Ball:2013hta}. The detector simulation for the hadron level events is performed using the fast detector simulator \texttt{Delphes-v3.5.0}~\cite{Selvaggi:2014mya}. The signal analyses at  LHC has been performed using \texttt{MadAnalysis-v5}~\cite{Conte:2012fm}.  
We generate the signal processes in \texttt{WHIZARD}~\cite{Kilian:2007gr} for the $e^+e^-$ and $\mu^+\mu^-$ collider studies. 
\subsection{At  HL-LHC} 
There are multiple possible final states which can probe the parameter space of our model. For our study, we consider the production of heavy Higgs ($h_3)$ via gluon fusion (GGF) leading to monojet+MET final state and vector boson fusion (VBF) production channels leading to two forward jets + MET final state at HL-LHC. Another important production mechanism of heavy Higgs can be via $b\bar b$ associated final state i.e. $b\bar{b}h_3$, which we postpone to a future study.

\begin{figure}
    \centering
    \begin{tikzpicture}
    \begin{feynman}
        \vertex (l1);
        \vertex[below right=of l1] (l3);
        \vertex[below left=of l3] (l2);
        \vertex[left=of l1] (i1) {\(g\)};
        \vertex[left=of l2] (i2) {\(g\)};
        \vertex[right=of l3] (f1) {\(h_3\)};
        \diagram* {(i1) --[gluon] (l1) --[fermion] (l3) --[fermion] (l2) --[fermion, edge label'=\(f\)] (l1), (i2) --[gluon] (l2), (l3) --[scalar] (f1),};
    \end{feynman}
    \end{tikzpicture}
    \begin{tikzpicture}
    \begin{feynman}
        \vertex (l1);
        \vertex[below right=of l1] (l2);
        \vertex[below left=of l2] (l3);
        \vertex[left=of l1] (i1) {\(q\)};
        \vertex[left=of l3] (i2) {\(q\)};
        \vertex[right=of l2] (f2) {\(h_3\)};
        \vertex[above=of f2] (f1) {\(q\)};
        \vertex[below=of f2] (f3) {\(q\)};
        \diagram* {(i1) --[fermion] (l1) --[boson, edge label'=\(V\)] (l2) --[boson, edge label'=\(V\)] (l3), (i2) --[fermion] (l3), (l2) --[scalar] (f2), (l1) --[fermion] (f1), (l3) --[fermion] (f3)};
    \end{feynman}
    \end{tikzpicture}
    \caption{Feynman diagrams contributing to gluon gluon fusion (GGF) and vector boson fusion (VBF) processes, created using Ref.~\cite{Feyn_diag:Ellis_2017}}
    \label{fig:feynlhc}
\end{figure}
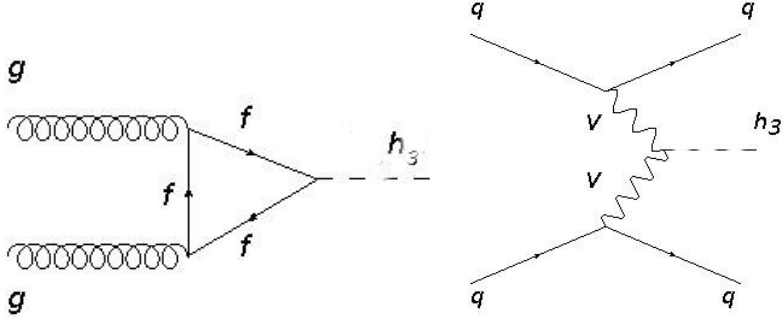

Fig.~\ref{fig:feynlhc} shows the  leading  Feynman diagrams for GGF and VBF processes.  
The  GGF  process is dominantly mediated by a top loop  and is followed by the decay of the heavy CP-even Higgs $h_3$ into a pair of DM candidates  manifested  as missing transverse energy.    Tagging the initial state radiation (ISR)  jet recoiling against the invisible system giving rise to mono-jet + $\slashed{E}_T$ signal leads to an observable final state  at colliders. 
The vector-boson fusion process (VBF) is characterized by two  jets widely separated in rapidity. An invisibly decaying heavy Higgs $h_3$ would lead to a final state consisting of two forward-moving jets along with missing transverse energy, i.e., 2 j + $\slashed{E}_T$.  
Therefore in order to look for the signal, we therefore consider the efficacy of the following   final states,
\begin{itemize}
\item Mono-jet + $\slashed{E}_T$,
\item 2 j + $\slashed{E}_T$ 
\end{itemize}
at the high luminosity LHC at $\sqrt{s}=14$ TeV and an integrated luminosity $\mathcal{L}=3000$ $fb^{-1}$. 
\subsubsection*{Signal Region A: Mono-jet + $\slashed{E}_T$}
We generated the gluon gluon fusion process with $h_3$ successively decaying into a pair of DM candidates in \texttt{Madgraph$\_$aMC$\_$v3.4.1}. For \textbf{BP1}, $\sigma_{GGF}\times BR(h_3 \times A_S A_S)$ = 0.232 fb. We perform the signal analyses using the following cuts successively from  Ref.~\cite{Dey:2019lyr} 
on the benchmark \textbf{BP1},
\begin{itemize}
\item \textbf{C1}: The  final state consists of  up to four jets with $p_T>30$ GeV and $|\eta|<2.8$.  
\item \textbf{C2}: We demand a large $\slashed{E}_T>250$ GeV.
\item \textbf{C3}: The hardest leading jet has $p_T>250$ GeV with $|\eta|<2.4$.
\item \textbf{C4}: We demand $\Delta \Phi ( j, \slashed{E}_T)>0.4 $ for all jets and  $\Delta \Phi ( j, \slashed{E}_T)>0.6$ for the leading jet. 
\item \textbf{C5}: A lepton-veto is imposed for electrons with $p_T>20$ GeV and $|\eta|<2.47$ and muons with $p_T>10$ GeV and $|\eta|<2.5$.
\end{itemize}
The SM background is obtained from the ATLAS mono-jet + $\slashed{E}_T$ search studied in Ref.~\cite{Dey:2019lyr}. 
\begin{table}[ht]
\begin{center}
\begin{tabular} {|c|c|c|c|c|c|}
\hline
Process & \textbf{C1} & \textbf{C2} & \textbf{C3} & \textbf{C4} & \textbf{C5} \\
 \hline
 GGF & 696& 137& 114&114 & 114\\
\hline 
 $\mathcal{S}$  & \multicolumn{5}{|c|}{ 1.356 $\sigma$} \\
 \hline 
\end{tabular}
\caption{The cut flow table for the number of signal events  for   \textbf{BP1} at leading order (LO) and signal significance $\mathcal{S}$ at $\sqrt{s}=14$ TeV and $\mathcal{L}=3000$ fb$^{-1}$. The SM background is obtained from Ref.~\cite{Dey:2019lyr}.}
\label{tab:GF}
\end{center}
\end{table} 
 We present the signal cut-flow table in Table.~\ref{tab:GF}. The statistical significance {(${\mathcal S}$)} of the signal  ($s$) over the total SM background ($b$) is calculated using
\cite{Cowan:2010js,ParticleDataGroup:2018ovx},
\begin{equation}
\mathcal{S} = \sqrt{2 \times \left[ (s+b){\rm ln}(1+\frac{s}{b})-s\right]},
\label{eq:sig}
\end{equation}
where $s$ and $b$ are the total signal and background event numbers after the cuts \textbf{C1-C5}. We observe that the GGF production process has a significance $\sim$ 1.36   $\sigma$ (LO)  at the  HL-LHC and is rather suppressed owing to the large mass of the heavy Higgs $m_{h_3}=900$ GeV leading to a low production cross-section. 

We normalize  the   GGF   production cross-section of $h_3$ using the K-factor computed from the gluon fusion cross-section for the 125 GeV Higgs. The gluon fusion cross-section computed using Madgraph for the  $125$ GeV Higgs at $\sqrt{s}=14$ TeV is 26.87 pb at leading order(LO) compared to  51.2 pb at NNLO+NNLL\cite{LHCHiggsCrossSectionWorkingGroup:2016ypw} resulting in a K-factor of 1.91. Using this K-factor, the signal significance improves to $\sim$2.6$\sigma$.
From the latest ATLAS analyses in the mono-jet + $\slashed{E}_T$ channel \cite{ATLAS:2021kxv}, a softer cut of 150 GeV is placed on the minimum transverse momentum of the leading jet and varying $250 \mbox{\rm}$  GeV$ \leq\slashed{E}_T\leq1200$ GeV. We obtain the best signal significance using this analysis for the $\slashed{E}_T>700$ GeV which results in the presence of $\sim 24 $ signal events, leading to an approximate  signal significance of $\sim  2.1 (3.5)  \sigma$ at LO (NNLO+NNLL). It is  assumed   a 10$\%$ increase in the $V$+jets background occurs when changing $\sqrt{s}$=13 TeV to 14 TeV, which contributes dominantly to the background. 
\subsubsection*{Signal Region B: 2 j + $\slashed{E}_T$}
Generating the  VBF process at LO analogously as described for the fusion process, we obtain, $\sigma_{VBF}\times BR(h_3 \rightarrow A_S A_S)$= 0.011 fb. We perform the signal analyses for the 2 j + $\slashed{E}_T$ final state,  using  the following cuts from \cite{Dey:2019lyr}  for the benchmark \textbf{BP1},
\begin{itemize}
\item \textbf{D1}: The final state consists of  at least two  jets with $p_T (j_1)>80$ and $p_T(j_2)>40$ GeV and $\Delta \Phi(j_i, \slashed{E}_T)>0.5$.
\item \textbf{D2}: We demand $\eta j_1j_2<0$ and $\Delta \Phi j_1 j_2 < 1.5$.
\item \textbf{D3}: We demand  $|\Delta \eta|_{jj}>3.0$.
\item \textbf{D4}: The invariant mass of the two forward jets is required to be large, i.e, $M_{jj}>600$ GeV.
\item \textbf{D5}: We demand  $\slashed{E}_T>200$ GeV.
\item \textbf{D6}: Furthermore, a lepton veto is imposed for   electrons with  $p_T >$ 20 GeV or muons with $p_T >$ 10
GeV.
\end{itemize}
\begin{table}[ht]
\begin{center}
\begin{tabular} {|c|c|c|c|c|c|c|}
\hline
Process & \textbf{D1} & \textbf{D2} & \textbf{D3} & \textbf{D4} & \textbf{D5}& \textbf{D6}\\
 \hline
VBF  & 1.25 &0.27 & 0.11& 0.11&0.11&0.11\\
\hline 
$\mathcal{S}$ & \multicolumn{6}{|c|}{ 0.0032  $\sigma$} \\
\hline 
\end{tabular}
\caption{The cut flow table for the  number of signal events 
for \textbf{BP1} at LO and signal significance $\mathcal{S}$  at $\sqrt{s}=14$ TeV and $\mathcal{L}=3000$ fb$^{-1}$. The SM background is obtained from Ref.~\cite{Dey:2019lyr}.}
\end{center}
\end{table} 
Rescaling the results with a K-factor (NNLO QCD+NLO) of 1.73 obtained from the ratio of the VBF production of the SM-like Higgs at NNLO QCD+NLO~\cite{LHCHiggsCrossSectionWorkingGroup:2016ypw} of 4.275 pb at $\sqrt{s}=14$ TeV compared to the production cross section 2.476 pb computed at LO in \texttt{Madgraph}, the signal significance improves to  0.0055 $\sigma$. We observe that due to the low production cross-section, the VBF channel is relatively more suppressed compared to the GGF channel and its observability is under doubt at the HL-LHC for \textbf{BP1}.  However, new machine learning techniques have been explored in Ref.~\cite{Dey:2019lyr} leading to an improvement in GGF and VBF channels in the context of the real singlet extension of 2HDM. Such techniques may also improve upon the signal in 2HDMS which we leave for future studies. 
\subsection{At Future Lepton Colliders}
We now discuss the prospects of  the 2HDMS  concerning the DM search, at proposed future lepton colliders such as $e^+e^-$ colliders ({\it eg.,} ILC \cite{Behnke:166034}, CLIC \cite{CLICdp:2018cto}) and  a muon collider~\cite{MuonCollider:2022nsa}. While at the LHC, GGF and VBF channels  give rise to the dominant contribution to the heavy Higgs production,  one has a better access to complementary processes such as mono-$X \ (X=\gamma,Z)$ + missing energy at lepton colliders, owing to a much cleaner environment compared to a hadron collider.  These final states can give rise to a clean channel for studying invisible Higgs decays into a pair of DM particles, with a visible particle ($X$) recoiling against the DM pair. For the current study, we present a signal-specific discussion of these processes at future lepton colliders and their comparison. We defer more detailed collider analyses for a future study.
\begin{figure}
    \centering
    \begin{tikzpicture}[scale=0.8, transform shape]
    \begin{feynman}[small]
        \vertex (b) ;
        \vertex[right=of b] (c);
        \vertex[above left=of b] (i3) {\(e^+/\mu^+\)};
        \vertex[below left=of b] (i4) {\(e^-/\mu^-\)};
        \vertex[above right=of c] (f3) {\(A_S\)};
        \vertex[below right=of c] (f4) {\(A_S\)};
        \diagram* {
        (i4) -- [fermion] (b) -- [fermion] (i3), 
        (b) -- [scalar, edge label'=\(h_{1,2,3}\)]  (c), (f3) -- [scalar] (c) -- [scalar] (f4),};
    \end{feynman}
    \end{tikzpicture}
    \begin{tikzpicture}[scale=0.8, transform shape]
    \begin{feynman}[small]
        \vertex (a) ;
        \vertex[above=of a] (b);
        \vertex[right=of b] (c);
        \vertex[above left=of b] (i3) {\(e^+/\mu^+\)};
        \vertex[below left=of a] (i4) {\(e^-/\mu^-\)};
        \vertex[above right=of c] (f3) {\(A_S\)};
        \vertex[below right=of c] (f4) {\(A_S\)};
        \vertex[below=of f4] (f5) {\(\gamma\)};
        \diagram* {
        (i4) -- [fermion] (a) --[fermion] (b) -- [fermion] (i3), 
        (b) -- [scalar, edge label'=\(h_{1,2,3}\)]  (c), (f3) -- [scalar] (c) -- [scalar] (f4), (a) --[photon] (f5)};
    \end{feynman}
    \end{tikzpicture}
    \caption{Feynman diagrams of $A_SA_S$ production process and the process with additional mono-photon at lepton colliders, created using Ref.~\cite{Feyn_diag:Ellis_2017}} 
    \label{fig:proc_xx}
\end{figure}
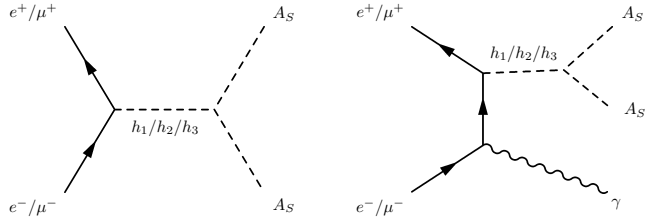
\noindent
The $A_SA_S\gamma$ final state can occur via the $s$ channel scalar-mediated $A_SA_S$ production with an initial state $\gamma$-radiation, as shown in Fig.~\ref{fig:proc_xx} (left). The same final state can also arise from the $t$-channel lepton-mediated process (as shown in  Fig.~\ref{fig:proc_xx} (right)). However, all these processes will be strongly suppressed by the small Yukawa couplings $c_{h_iee}$ at the $e^+ e^-$ collider, while the muon collider can have a sizeable production cross-section due to the larger $c_{h_i\mu\mu}$ couplings. 
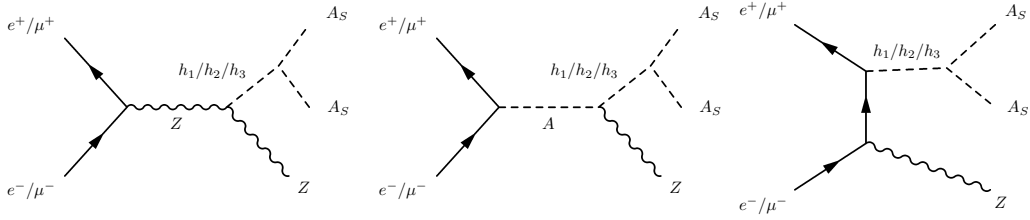
\begin{figure}
    \centering
    \begin{tikzpicture}[scale=0.8, transform shape]
    \begin{feynman}[small]
        \vertex (b) ;
        \vertex[right=of b] (c);
        \vertex[above left=of b] (i3) {\(e^+/\mu^+\)};
        \vertex[below left=of b] (i4) {\(e^-/\mu^-\)};
        \vertex[above right=of c] (d);
        \vertex[above right=of d] (f3) {\(A_S\)};
        \vertex[below right=of d] (f4) {\(A_S\)};
        \vertex[below=of f4] (f5) {\(Z\)};
        \diagram* {
        (i4) --[fermion] (b) --[fermion] (i3), 
        (b) --[boson, edge label'=\(Z\)]  (c), (c) --[scalar, edge label'=\(h_{1,2,3}\)] (d), (f3) --[scalar] (d) --[scalar] (f4), (c) --[boson] (f5)};
    \end{feynman}
    \end{tikzpicture}
    \begin{tikzpicture}[scale=0.8, transform shape]
    \begin{feynman}[small]
        \vertex (b) ;
        \vertex[right=of b] (c);
        \vertex[above left=of b] (i3) {\(e^+/\mu^+\)};
        \vertex[below left=of b] (i4) {\(e^-/\mu^-\)};
        \vertex[above right=of c] (d);
        \vertex[above right=of d] (f3) {\(A_S\)};
        \vertex[below right=of d] (f4) {\(A_S\)};
        \vertex[below=of f4] (f5) {\(Z\)};
        \diagram* {
        (i4) --[fermion] (b) --[fermion] (i3), 
        (b) --[scalar, edge label'=\(A\)]  (c), (c) --[scalar, edge label'=\(h_{1,2,3}\)] (d), (f3) --[scalar] (d) --[scalar] (f4), (c) --[boson] (f5)};
    \end{feynman}
    \end{tikzpicture}
    \begin{tikzpicture}[scale=0.8, transform shape]
    \begin{feynman}[small]
        \vertex (a) ;
        \vertex[above=of a] (b);
        \vertex[right=of b] (c);
        \vertex[above left=of b] (i3) {\(e^+/\mu^+\)};
        \vertex[below left=of a] (i4) {\(e^-/\mu^-\)};
        \vertex[above right=of c] (f3) {\(A_S\)};
        \vertex[below right=of c] (f4) {\(A_S\)};
        \vertex[below=of f4] (f5) {\(Z\)};
        \diagram* {
        (i4) -- [fermion] (a) --[fermion] (b) -- [fermion] (i3), 
        (b) -- [scalar, edge label'=\(h_{1,2,3}\)]  (c), (f3) -- [scalar] (c) -- [scalar] (f4), (a) --[photon] (f5)};
    \end{feynman}
    \end{tikzpicture}
    \caption{Feynman diagrams of $Z A_SA_S$ production process at lepton colliders, created using Ref.~\cite{Feyn_diag:Ellis_2017}}
    \label{fig:proc_zxx}
\end{figure}
\begin{figure}
    \centering
    \begin{tikzpicture}[scale=0.8, transform shape]
    \begin{feynman}[small]
        \vertex (b) ;
        \vertex[right=of b] (c);
        \vertex[below left=of b] (a);
        \vertex[below left=of a] (i4) {\(e^-/\mu^-\)};
        \vertex[below right=of a] (f1) {\(\gamma\)};
        \vertex[above left=of b] (b2);
        \vertex[above left=of b2] (i3) {\(e^+/\mu^+\)};
        \vertex[above right=of c] (d);
        \vertex[above right=of d] (f3) {\(A_S\)};
        \vertex[below right=of d] (f4) {\(A_S\)};
        \vertex[below=of f4] (f5) {\(Z\)};
        \diagram* {
        (i4) --[fermion] (a) --[fermion] (b) --[fermion] (i3), 
        (b) --[boson, edge label'=\(Z\)]  (c), (c) --[scalar, edge label'=\(h_{1,2,3}\)] (d), (f3) --[scalar] (d) --[scalar] (f4), (c) --[boson] (f5), (a) --[photon] (f1)};
    \end{feynman}
    \end{tikzpicture}
    \begin{tikzpicture}[scale=0.8, transform shape]
    \begin{feynman}[small]
        \vertex (b) ;
        \vertex[right=of b] (c);
        \vertex[below left=of b] (a);
        \vertex[below left=of a] (i4) {\(e^-/\mu^-\)};
        \vertex[below right=of a] (f1) {\(\gamma\)};
        \vertex[above left=of b] (b2);
        \vertex[above left=of b2] (i3) {\(e^+/\mu^+\)};
        \vertex[above right=of c] (d);
        \vertex[above right=of d] (f3) {\(A_S\)};
        \vertex[below right=of d] (f4) {\(A_S\)};
        \vertex[below=of f4] (f5) {\(Z\)};
        \diagram* {
        (i4) --[fermion] (a) --[fermion] (b) --[fermion] (i3), 
        (b) --[scalar, edge label'=\(A\)]  (c), (c) --[scalar, edge label'=\(h_{1,2,3}\)] (d), (f3) --[scalar] (d) --[scalar] (f4), (c) --[boson] (f5), (a) --[photon] (f1)};
    \end{feynman}
    \end{tikzpicture}
    \begin{tikzpicture}[scale=0.8, transform shape]
    \begin{feynman}[small]
        \vertex (a) ;
        \vertex[above=of a] (b);
        \vertex[above=of b] (b2);
        \vertex[right=of b2] (c);
        \vertex[above left=of b2] (i3) {\(e^+/\mu^+\)};
        \vertex[below left=of a] (i4) {\(e^-/\mu^-\)};
        \vertex[above right=of c] (f3) {\(A_S\)};
        \vertex[below right=of c] (f4) {\(A_S\)};
        \vertex[below=of f4] (f5) {\(Z/\gamma\)};
        \vertex[below=of f5] (f6) {\(Z/\gamma\)};
        \diagram* {
        (i4) -- [fermion] (a) --[fermion] (b) --[fermion] (b2) -- [fermion] (i3), 
        (b2) -- [scalar, edge label'=\(h_{1,2,3}\)]  (c), (f3) -- [scalar] (c) -- [scalar] (f4), (b) --[boson] (f5), (a) --[boson] (f6)};
    \end{feynman}
    \end{tikzpicture}
    \caption{Feynman diagrams of $Z A_SA_S \gamma$ production process at lepton colliders, created using Ref.~\cite{Feyn_diag:Ellis_2017}}
    \label{fig:proc_zxxisr}
\end{figure}
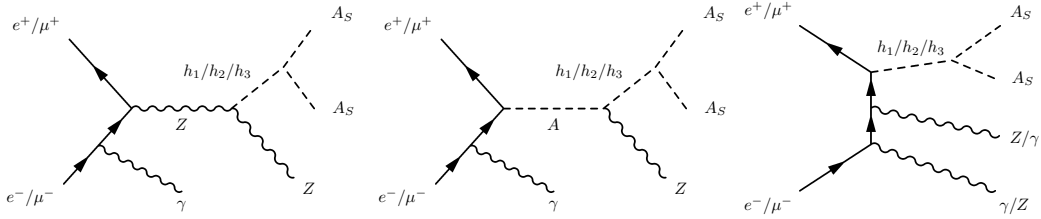

\noindent
Fig.~\ref{fig:proc_zxx} shows the Feynman diagrams corresponding to $ZA_SA_S$ final state. The DM candidate $A_S$ can be produced by the $h_iA_SA_S$ interaction, while in the lepton colliders Higgs bosons can be produced via the Higgsstrahlung process (left diagram) and yield the $ZA_SA_S$ final state. In addition, the processes involving the Yukawa couplings, shown as in  the  center and right diagrams of Fig.~\ref{fig:proc_zxx}, can also give rise to the $ZA_SA_S$ final state. In Fig.~\ref{fig:proc_zxxisr},  we show the diagrams of $ZA_SA_S$ processes with an additional photon. The processes involving Yukawa couplings will suffer significant suppression at the $e^+e^-$-collider for similar reasons as discussed for the $A_SA_S\gamma$ final state.
\begin{figure}[ht]
   \centering
   \includegraphics[scale=0.45]{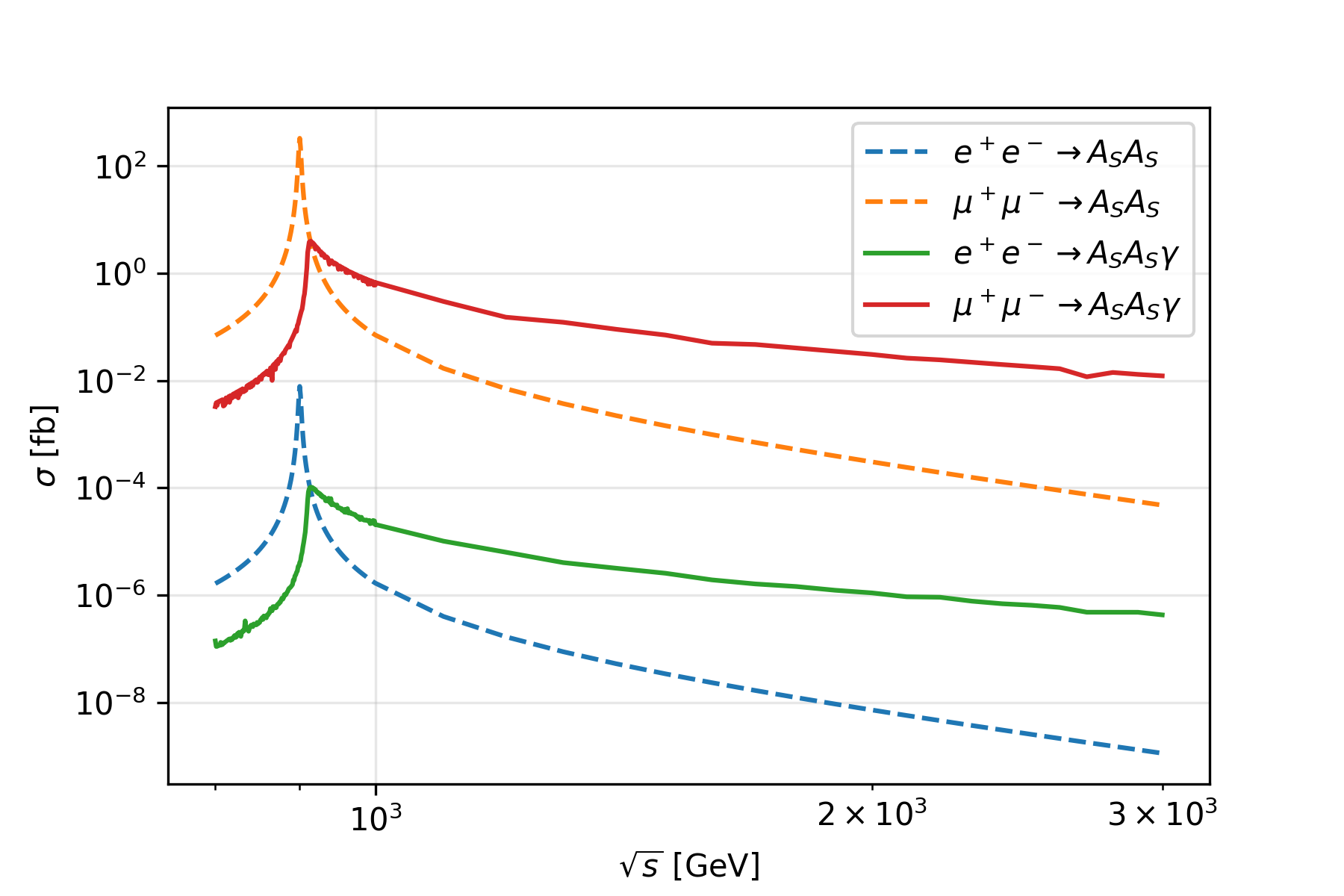}
   \includegraphics[scale=0.45]{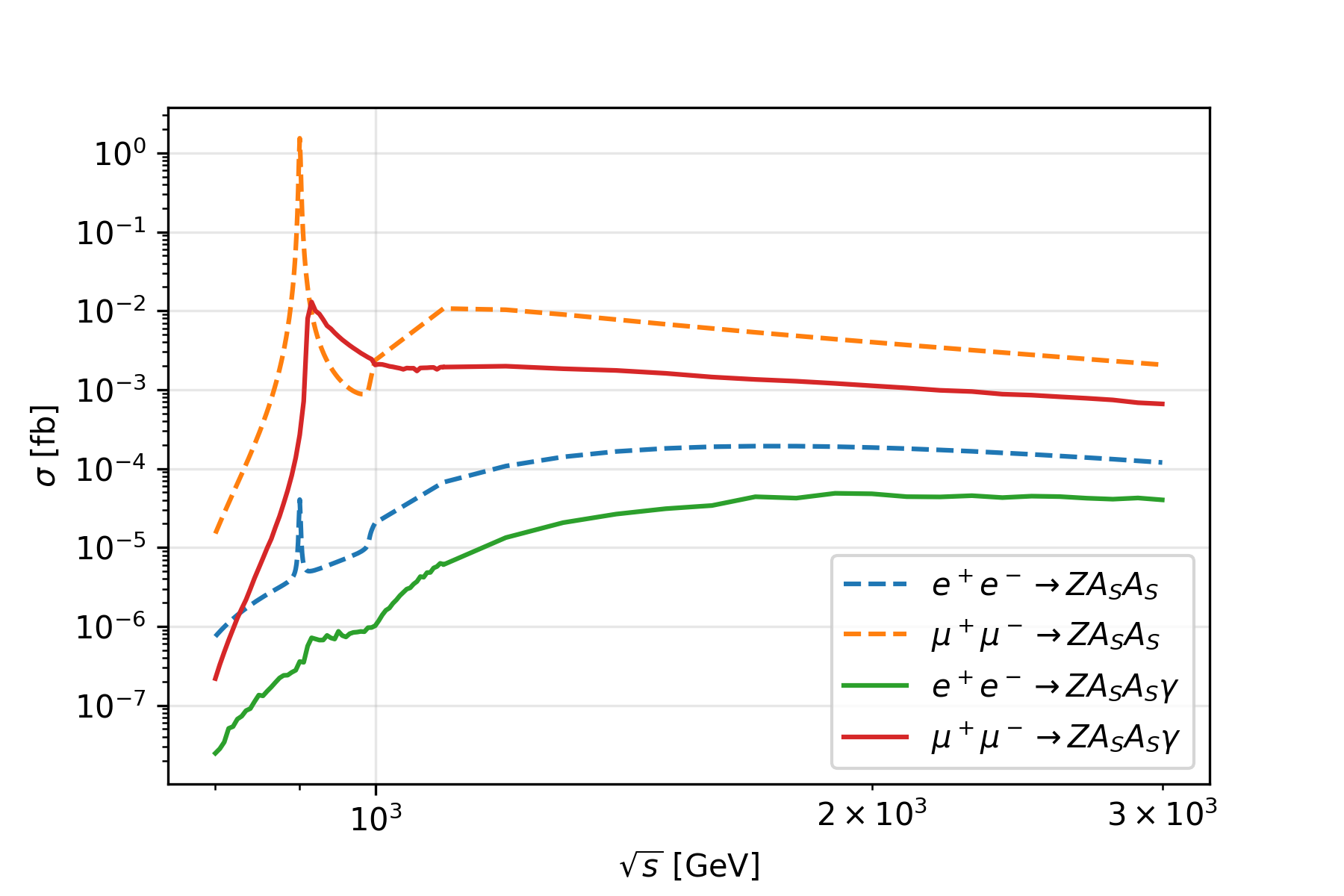}
    \caption{Variation of the cross-section vs. $\sqrt{s}$ at $e^+e^-$ and muon collider for the processes $A_SA_S$ (dotted) and $A_SA_S\gamma$ (solid) in the left panel, and $ZA_SA_S$ (dotted) and $ZA_SA_S\gamma$ (solid) final states on the right panel.}
    \label{fig:emu} 
\end{figure}

\noindent
In Fig.~\ref{fig:emu} (left), we present the cross-sections for $A_SA_S\gamma$ final state (solid curves) as a function of $\sqrt{s}$. We have  also  shown the cross-section of the process $e^+e^-/\mu^+\mu^- \rightarrow A_SA_S$ (dotted curves)   in the same plot for understanding. Although all the scalars ($h_1,h_2,h_3$) take part in the aforementioned process, the major contribution comes from the resonant-$h_3$ production, around $\sqrt{s}=900$ GeV. Before this resonance, the off-shell production of $h_3$ and its subsequent decay to $A_SA_S$ dominates, where $\gamma$ is radiated off the initial state leptons. Evidently, in this region the cross-section of $e^+e^-/\mu^+\mu^- \rightarrow A_SA_S\gamma$ is suppressed compared to the cross-section of  $e^+e^-/\mu^+\mu^- \rightarrow A_SA_S$. We see that at $h_3$-resonance at $\sqrt{s}=900$ GeV, all the cross-sections reach their maxima. Beyond the resonance, the $e^+e^-/\mu^+\mu^- \rightarrow A_SA_S\gamma$ is dominated by the $t$-channel production of on-shell $h_3$ and its subsequent decay into a DM pair. Since $e^+e^-/\mu^+\mu^- \rightarrow A_SA_S$ takes place solely via the $s$-channel mediation, its cross-section falls off beyond the resonance. Therefore in this region, $e^+e^-/\mu^+\mu^- \rightarrow A_SA_S\gamma$ shows an enhancement over $e^+e^-/\mu^+\mu^- \rightarrow A_SA_S$ cross-section. All these processes acquire an enhancement of around 5 orders of magnitude in the muon collider as compared to the electron-positron collider, due to Yukawa enhancement. \\
We also consider $ZA_SA_S$ and $ZA_SA_S$ along with a photon which would also lead to clean final states at lepton colliders. Fig.~\ref{fig:emu} (right) shows the corresponding production cross-sections.  The $ZA_SA_S$ as well as $ZA_SA_S\gamma$ processes have an enhancement at the $h_3$-resonance. Before the resonance, the Higgsstrahlung process as well as the off-shell scalar mediated processes make contributions to the final states. Beyond the resonance, the $t$-channel production of on-shell $h_3$ and $Z$ processes dominates. 
Similar to the mono-photon case, here too, all the distributions follow a similar shape for the $e^+e^-$ and muon colliders. We also see an overall enhancement of the cross-section by two-to-five orders of magnitude at the muon   collider as compared to the $e^+e^-$ collider. \\
Concerning  all the processes involving photons in the final state, we employ the following cuts on the photon:  $E_{\gamma} > 10$ GeV and $\theta>7^\circ$ \cite{Kalinowski:2020lhp} 
during event generation in \texttt{WHIZARD}, in order to avoid divergences, especially in case of ISR photons. \\
In principle, a $\mu^+\mu^-$ collider has a better prospect of observing these  processes and the $\mu^+\mu^- \rightarrow A_SA_S\gamma$ offers the largest cross-section amongst all the aforementioned processes. In the present study we provide estimates of production cross-sections for a benchmark signal process. Although we do not perform background analysis, we generally expect a cleaner environment of the lepton collider compared to the hadron collider. One should also keep in mind that the advantage of initial beam polarization can further help reduce the background and/or enhance the signal. A detailed study  on this topic, we reserve for a future study.  
\section{Summary and Conclusions}
\label{sec:summ}
In this work, we focus on the 2HDM + a complex singlet scalar $S$, under the assumption, that the complex singlet is odd under a $Z_2$ symmetry. We further assume, that the imaginary part of the complex singlet does not get a {\it vev}, but the real part  acquires a {\it vev}, giving rise to a mixing between the singlet and the scalar sector of the 2HDM. Such a mixing between the two sectors is an important feature of the model, which motivates us to look for the possibility of embedding a 95 GeV scalar as well. The recent excess at the CMS experiments in the $\gamma\gamma(2.8\sigma$) and $\tau\tau(2.6\sigma)$ final states, as well as the LEP excess around the similar mass range in the $b\bar b(2.3\sigma)$ final state can thus be explained in significant regions of our model parameter space. We have performed a thorough scan and identified those regions that are allowed by all the existing constraints, namely the theoretical constraints such as bfb, tree-level  unitarity as well as experimental constraints from direct search and precision observables.   we focused on the observed excesses at the CMS and LEP experiments and identified the region that is  consistent with the observed signal strength($\mu$) of the 95 GeV state. \\
Although the $Z_2$ symmetry is spontaneously broken by the {\it vev} of the real part of the complex singlet, the zero {\it vev} condition of the imaginary part makes it a viable DM candidate ($A_S$). This is another crucial aspect of this model. It can not only provide a plausible explanation for the observed excess, but can also accommodate a suitable DM candidate. We have checked all the DM constraints, namely, the observed relic density,  the direct  and indirect detection bounds, and presented the allowed parameter space. Finally, we chose a suitable benchmark, which is allowed by all the aforementioned constraints and fits the observed excess. Thereafter, we explored the detection possibility of the benchmark at the high-luminosity LHC and future lepton colliders. In the collider search, our main focus is on DM phenomenology. Therefore, we look for mono-jet and two forward jet + $\slashed{E_T}$ final state at the high-luminosity LHC. We found out that for our given benchmark, the HL-LHC projections are not promising, owing to the high masses of the heavy scalars. However, we emphasize that the benchmark is chosen for illustration purpose and in principle, a benchmark with a lower non-standard scalar masses are possible, which may be probed at the HL-LHC. Also, there is a promising potential of improvement with machine-learning techniques, that we do not consider in this work. Instead, we focused on a complimentary search strategy, at the lepton colliders and looked at mono-photon or mono-Z final states. We found that the muon collider shows the best prospect in terms of production cross-sections. A detailed background analysis and also possible improvement with beam polarization at the lepton colliders, is postponed for a future study. 
\section*{Acknowledgements}
 JD, JL, GMP and JZ acknowledge support by the Deutsche Forschungsgemeinschaft (DFG, German Research Foundation) under Germany's Excellence Strategy EXC 2121 "Quantum Universe"- 390833306.  JD acknowledges support from the  HEP  Dodge Family Endowment Fellowship at the Homer L.Dodge Department of Physics $\&$ Astronomy at the University of Oklahoma.
\providecommand{\href}[2]{#2}
\bibliographystyle{JHEP}   
\bibliography{ref}

\appendix
\section{Feynman Diagrams}
\begin{figure}[h!]
\centering
\begin{tikzpicture}
\begin{feynman}[small]
\vertex (li) {\(A_S\)};
\vertex [below=2cm of li] (hi) {\(N\)};
\vertex [right=of li] (a);
\vertex [above right=of a] (lf) {\(A_S\)};
\vertex [below right=of a] (b);
\vertex [right=of b] (hf1);
\vertex [blob, right=of hi] (c) {};
\path (c.-10) ++ (00:2) node[vertex] (hf2);
\path (c.-40-|hf2.center) node[vertex] (hf3);
\diagram* {
    (li) -- [scalar] (a) -- [scalar] (lf),
    (hi) -- [fermion] (c) -- [fermion] (b),
    (a) -- [scalar, edge label'=\(h_j\)] (b) -- [fermion] (hf1),
    (c.-10) -- [fermion] (hf2),
    (c.-40) -- [fermion] (hf3)};
\end{feynman}
\end{tikzpicture}
\caption{Feynman diagram relevant for calculation of direct detection DM-proton/neutron cross-section, created using Ref.~\cite{Feyn_diag:Ellis_2017}}
\label{fig:feynman_diag_cs}
\end{figure}

\begin{figure}[h!]
\centering
\begin{tikzpicture}
\begin{feynman}[small]
\vertex (a);
\vertex[above left=of a] (i1) {\(A_S\)};
\vertex[below left=of a] (i2) {\(A_S\)};
\vertex[above right=of a] (f1) {\(h_j\)};
\vertex[below right=of a] (f2) {\(h_k\)};
\diagram* {(i1) --[scalar] (a) --[scalar] (i2), (a) --[scalar] (f1), (a) --[scalar] (f2),};
\end{feynman}
\end{tikzpicture}
\begin{tikzpicture}
\begin{feynman}[small]
\vertex (b) ;
\vertex[right=of b] (c);
\vertex[above left=of b] (i3) {\(A_S\)};
\vertex[below left=of b] (i4) {\(A_S\)};
\vertex[above right=of c] (f3) {\(h_k\)};
\vertex[below right=of c] (f4) {\(h_l\)};
\diagram* {
(i3) -- [scalar] (b) -- [scalar] (i4), 
(b) -- [scalar, edge label'=\(h_j\)]  (c), (f3) -- [scalar] (c) -- [scalar] (f4),};
\end{feynman}
\end{tikzpicture}
\begin{tikzpicture}
\begin{feynman}[small]
\vertex (b) ;
\vertex[right=of b] (c);
\vertex[above left=of b] (i3) {\(A_S\)};
\vertex[below left=of b] (i4) {\(A_S\)};
\vertex[above right=of c] (f3) {\(SM\)};
\vertex[below right=of c] (f4) {\(SM\)};
\diagram* {
(i3) -- [scalar] (b) -- [scalar] (i4), 
(b) -- [scalar, edge label'=\(h_j\)]  (c), (f3) -- [boson] (c) -- [boson] (f4),};
\end{feynman}
\end{tikzpicture}
\begin{tikzpicture}
\begin{feynman}[small]
\vertex (b) ;
\vertex[right=of b] (c);
\vertex[above left=of b] (i3) {\(A_S\)};
\vertex[below left=of b] (i4) {\(A_S\)};
\vertex[above right=of c] (f3) {\(SM\)};
\vertex[below right=of c] (f4) {\(SM\)};
\diagram* {
(i3) -- [scalar] (b) -- [scalar] (i4), 
(b) -- [scalar, edge label'=\(h_j\)]  (c), (f3) -- [anti fermion] (c) -- [fermion] (f4),};
\end{feynman}
\end{tikzpicture}
\begin{tikzpicture}
\begin{feynman}[small]
\vertex (a);
\vertex[below=of a] (b);
\vertex[above left=of a] (i1) {\(A_S\)};
\vertex[below left=of b] (i2) {\(A_S\)};
\vertex[above right=of a] (f1) {\(h_j\)};
\vertex[below right=of b] (f2) {\(h_k\)};
\diagram* {(i1) --[scalar] (a) --[scalar, edge label'=\(A_S\)] (b) --[scalar] (i2), (a) --[scalar] (f1), (b) --[scalar] (f2),};
\end{feynman}
\end{tikzpicture}
\begin{tikzpicture}
\begin{feynman}[small]
\vertex (a);
\vertex[below=of a] (b);
\vertex[above left=of a] (i1) {\(A_S\)};
\vertex[below left=of b] (i2) {\(A_S\)};
\vertex[below right=of b] (f1) {\(h_j\)};
\vertex[above right=of a] (f2) {\(h_k\)};
\diagram* {(i1) --[scalar] (a) --[scalar, edge label'=\(A_S\)] (b) --[scalar] (i2), (a) --[scalar] (f1), (b) --[scalar] (f2),};
\end{feynman}
\end{tikzpicture}
\caption{Tree level Feynman diagrams relevant for calculation of relic density and indirect detection annihilation cross-section, created using Ref.~\cite{Feyn_diag:Ellis_2017}}
\label{fig:feynman_diag_relden}
\end{figure}

\end{document}